\documentclass[12pt]{article}
\usepackage[utf8]{inputenc}
\usepackage{subfig}
\usepackage{tikz}  
\usetikzlibrary{shapes,arrows}

\usepackage{setspace}
\usepackage[graphicx]{realboxes}
\usepackage[font=onehalfspacing]{caption}
\usepackage{adjustbox}
\usepackage{multirow,array}
\usepackage{natbib}
\usepackage{bm}
\usepackage{bbm}

\newcolumntype{P}[1]{>{\centering\arraybackslash}p{#1}}

\usepackage{blindtext}
\usepackage{fancyhdr}

\usepackage[margin=1in]{geometry}

\newcommand{\amend}[1]{{\color{black} #1}}

\newcommand{\beginsupplement}{%
        \setcounter{table}{0}
        \renewcommand{\thetable}{S\arabic{table}}%
        \setcounter{figure}{0}
        \renewcommand{\thefigure}{S\arabic{figure}}%
     }

\usepackage[utf8]{inputenc} 
\usepackage[T1]{fontenc}    
\usepackage{hyperref}       
\usepackage{url}            
\usepackage{booktabs}       
\usepackage{amsfonts}       
\usepackage{nicefrac}       
\usepackage{microtype}      
\usepackage{graphicx}
\usepackage{amsmath}
\usepackage{booktabs}
\usepackage{verbatim}
\usepackage{mathtools}

\usepackage{color}

\begin{document}

\fancyhead[L]{hi}
\fancyhead[R]{\today}

\title{NFL Ghosts: A framework for evaluating defender positioning with conditional density estimation
}
\author{Ronald Yurko$^\text{1}$ \and Quang Nguyen$^\text{1}$ \and Konstantinos Pelechrinis$^\text{2}$}
\date{%
    \normalsize
    $^\text{1}$Department of Statistics \& Data Science, Carnegie Mellon University\\%
    $^\text{2}$Department of Informatics and Networked Systems, University of Pittsburgh \\[2ex]
}

\maketitle

\begin{abstract}

Player attribution in American football remains an open problem due to the complex nature of twenty-two players interacting on the field, but the granularity of player tracking data provides ample opportunity for novel approaches. In this work, we introduce the first public framework to evaluate spatial and trajectory tracking data of players relative to a baseline distribution of ``ghost'' defenders. We demonstrate our framework in the context of modeling the nearest defender positioning at the moment of catch. In particular, we provide estimates of how much better or worse their observed positioning and trajectory compared to the expected play value of ghost defenders. Our framework leverages \amend{multi}-dimensional tracking data features through flexible random forests for conditional density estimation in two ways: (1) to model the distribution of receiver yards gained enabling the estimation of within-play expected value, and (2) to model the 2D spatial distribution of baseline ghost defenders. We present novel metrics for measuring player and team performance based on tracking data, and discuss challenges that remain in extending our framework to other aspects of American football.

\end{abstract}


\newpage

\section{Introduction}
\label{sec:intro}

Player tracking data have become a powerful resource in sports analytics in recent years.
Many professional sports leagues around the world collect data on spatial coordinates of players and ball, enabling continuous-time modeling and analysis of sports at a fine-grained level.
This is a massive leap forward from traditional data sources like box-score statistics and event-by-event data.
For recent surveys on player tracking data in sports, we refer the reader to \cite{baumer2023big} and \cite{kovalchik2023player}.

Player tracking data provide a great opportunity for novel advancements in statistical modeling and analysis of sports.
In this paper, we focus on one specific task in sports analytics: player evaluation.
Specifically, we propose a framework for evaluating players based on the rich tracking data features on player positioning and trajectory.
We illustrate our framework with American football, as we aim to provide an assessment of defensive pass coverage performance in the context of limiting yards after catch.
Below, we briefly highlight notable contributions on play evaluation, tracking data models, and player evaluation in sports.

\subsection{Previous work: expected value of a game state}



The task of estimating the expected value of a game state given its contextual information is fundamental in sports analytics.
Prior to tracking data became available, this task is performed at the discrete level, where a sport can be divided into categorically distinct events or actions.
In baseball, the runs expectancy matrix \citep{lindsey1963investigation} contains the expected number of runs in the remainder of an inning for each situation of number of outs and runners on base.
In soccer, expected goals (xG) \citep{barnett1993effect} is commonly used to value a shot attempt.
xG measures the conversion probability of a shot by accounting for various factors, including distance and angle to goal, part of the body used for shooting the ball, and type of previous event.

In American football, expected points (EP) is a common concept for estimating the expected value of a game state.
The game state features in an EP model typically include the current down, yards to go for a first down, current yardline, time remaining, to name a few.
Importantly, EP has the advantage of interpretability.
Since its unit of measurement is on a point scale, EP provides are useful measure for in-game strategic decision making \citep{romer2006firms}.
For example, one can take the difference between pre-play EP and post-play EP to measure the value for an individual play.
This is commonly known as expected points added (EPA) \citep{burke2009expected, yurko2019nflwar}. \amend{EPA values can range from roughly -14 to +14, with the endpoints reflecting extreme swings in the game state value (e.g., a turnover by the offense on their opponent's goal line that leads to a defensive touchdown).}

Early attempts on modeling EP in football provide estimates at the discrete-time between-play level.
\cite{yurko2019nflwar} summarize the key points and limitations of previous work \citep{carter1971operations, carroll1988hidden}, before proposing a multinomial logistic regression model to estimate EP.
This approach first estimates the probability of the next scoring outcome, and then computes EP as a function of these probabilities.
The considered scoring outcomes and their respective point values are touchdown (7), field goal (3), safety (2), no score (0), opposing team safety (-2), opposing team field goal (-3), and opposing team touchdown (-7).

At the continuous level, the granularity of player tracking data allows for the estimation of within-play expected values.
\cite{cervone2016multiresolution} propose a continuous-time framework for estimating the expected possession value (EPV) for the offensive team during a possession in basketball.
This framework consists of two sub-models, one for player movement and one for player decision-making outcome (pass, shot, or turnover).
\cite{sicilia2019deephoops} introduce a different in-play expected value approach for basketball.
This work first estimates the probability for within-play actions, namely, shot attempt, foul (shooting and non-shooting), and turnover, before converting to a point scale.
In American football, \cite{yurko2020going} provide continuous-time within-play valuation estimates by using a long short-term memory model to predict the expected yards gained for a ball-carrier throughout a play.
\cite{fernandez2021framework} use deep learning to develop EPV for soccer, based on the values of three main actions: shot, pass, and drive.
For more discussions on within-play expected value in sports, see \cite{macdonald2020recreating}.

\subsection{Previous work: player tracking data in American football}

In American football, the National Football League (NFL) collects tracking data via their tracking system known as Next Gen Stats by having radio frequency identification (RFID) tags installed into players' shoulder pads and the football.
These high-resolution data, measured at a rate of 10 frames per second, provide positional information for all players on the field (and also the ball).
Along with their player tracking system, the NFL also launched the Big Data Bowl---an annual data science competition focusing on player tracking data---in 2018 \citep{nfl2024big}.
Each year, the NFL Big Data Bowl releases a sample of tracking data to accompany a competition theme on a specific aspect of American football such as pass coverage, special teams, and linemen on pass plays.
The competition has paved the path for many innovations in football analytics and subsequently established the foundation for the peer-reviewed literature on statistical analysis of football tracking data \citep{lopez2020bigger}.



\cite{chu2020route} implement model-based clustering to characterize receiver routes on passing plays.
\cite{yurko2020going} introduce a framework for obtaining continuous-time estimates of play value in American football.
\cite{deshpande2020expected} propose an expected hypothetical completion probability framework for assessing offensive plays and quarterback decision making.
\cite{dutta2020unsupervised} use unsupervised learning methods to identify pass coverage types for NFL defensive backs.
\cite{nguyen2024here} and \cite{nguyen2025fractional} devise novel metrics for evaluating pass rush and tackling in American football at the frame level.
Note that these articles provide novel means for assessing both offensive and defensive players in American football.
These are remarkable accomplishments, since there were no comprehensive measures for evaluating positions such as defensive linemen prior to player tracking data \amend{becoming} publicly available \citep{wolfson2017forecasting}.

Along with the contribution to their respective evaluation aspect of football, the aforementioned work all share one common theme.
That is, their models consist of features obtained from tracking data based on player positioning and trajectory.
As classified by \cite{kovalchik2023player}, there are two main types of features for measuring player performance: directly-derived features and model-based features.
Directly-computed variables such as distance, velocity, acceleration, orientation and angle for each player (and between players in the case of distance and angle) are quintessential to developments in player evaluation with tracking data.
For instance, \cite{nguyen2024here} propose a model-free, continuous pass rush metric by simply taking the ratio of the pass rusher's velocity toward the quarterback and distance to the quarterback.
As for model-based features, one common strategy is to define space ownership for each player on the field.
For example, \cite{yurko2020going} construct different features with respect to football field region owned by the ball carrier obtained from Voronoi tessellation.

Nevertheless, the task of player evaluation \amend{based on tracking data features that are functions of the players themselves (e.g., location, speed, direction)} remains an open problem.


\subsection{Previous work: player evaluation with ghosting}




Ghosting is a prominent approach for evaluating players in continuous-time based on player positioning and movement information.
As its name would suggest, ghosting aims to identify optimal player actions throughout a play by comparing a player to an average player, i.e., ``ghost''.
This notion is first introduced in basketball, as \cite{lowe2013lights} illustrates how the Toronto Raptors at the time were using tracking data to perform player evaluation.
In particular, the team develops an algorithm for comparing players in reality to ghost players, in order to analyze the actual player actions on the basketball court versus what they should have done at any particular moment within a possession.

Perhaps the most common data-driven technique for ghosting is \textit{deep imitation learning}, which is first proposed by \cite{le2017data}.
This seminal work examines defensive strategies in soccer, studying where a defender at any given moment within a play should have been based on a league average model.
Here, the optimal locations are those that minimize the offense's chance of scoring given positional information of every player on the pitch.
\cite{le2017coordinated} then follow-up by proposing a coordinated multi-agent imitation learning framework in soccer.
This strategy uses a long short-term memory network to model player trajectories and a three-phase training process that alternates between evaluating individual player policies and a team's joint policy.
\cite{le2017coordinated}'s coordinated multi-agent model is then adopted in basketball and American football.
In basketball, \cite{seidl2018bhostgusters} propose an interactive player sketching system, which outputs ghost players that imitate basketball defensive behaviors after receiving information about offensive trajectories.
In American football, \cite{schmid2021simulating} offer a ghosting framework which generates defensive trajectories and allows for the comparison of player movement within a play with a simulated league-average behavior.
The article also introduces a pass completion probability model to evaluate the proposed ghosting model for understanding defensive behavior in the NFL.

Note that the aforementioned work provide ghosting models for only the defensive players and team. 
\cite{felsen2018where} use conditional variational autoencoders to simulate the offense in basketball and predict the personalized adversarial motion of players on the court.
Indeed, recent developments of ghosting have seen the rise of \textit{generative models} in their methodology.
\cite{zhan2019generating} propose a method for simulating how a basketball team would execute a play given a defined ``macro intent'' (e.g., setting up a formation).
This allows for the understanding of player trajectories and how they interact with one another to reach a specific end state and achieve a goal.
\cite{gu2023deep} introduce a deep generative ghosting model using conditional variational recurrent neural network to imitate player movement and interactions in soccer.
This approach first creates sequences of pitch control grids to represent player interactions, before training the generative model to generate benchmark player and team performances.
More recently, \cite{srinivasan2023thinking} combine imitation learning with generative and language models to imitate the playing style and shot selection of tennis players.


\subsection{Our contributions}

In this paper, we propose a framework for evaluating player positioning and trajectory relative to baseline ghost players.
In doing so, we use a \amend{multi}-dimensional conditional density estimation approach to account for a rich set of features derived from tracking data to compute the expected within-play value.
Our flexible framework provides evaluation on the scale of expected points, thus having the benefit of interpretability.
Further, the expected within-play value for ghosts can be computed using a full probability distribution of player positioning and movement information.
This overcomes a major limitation of previous ghosting approaches, which only return a point estimate for the optimal player location and trajectory.

\begin{figure}
    \centering
    \includegraphics[scale=0.6]{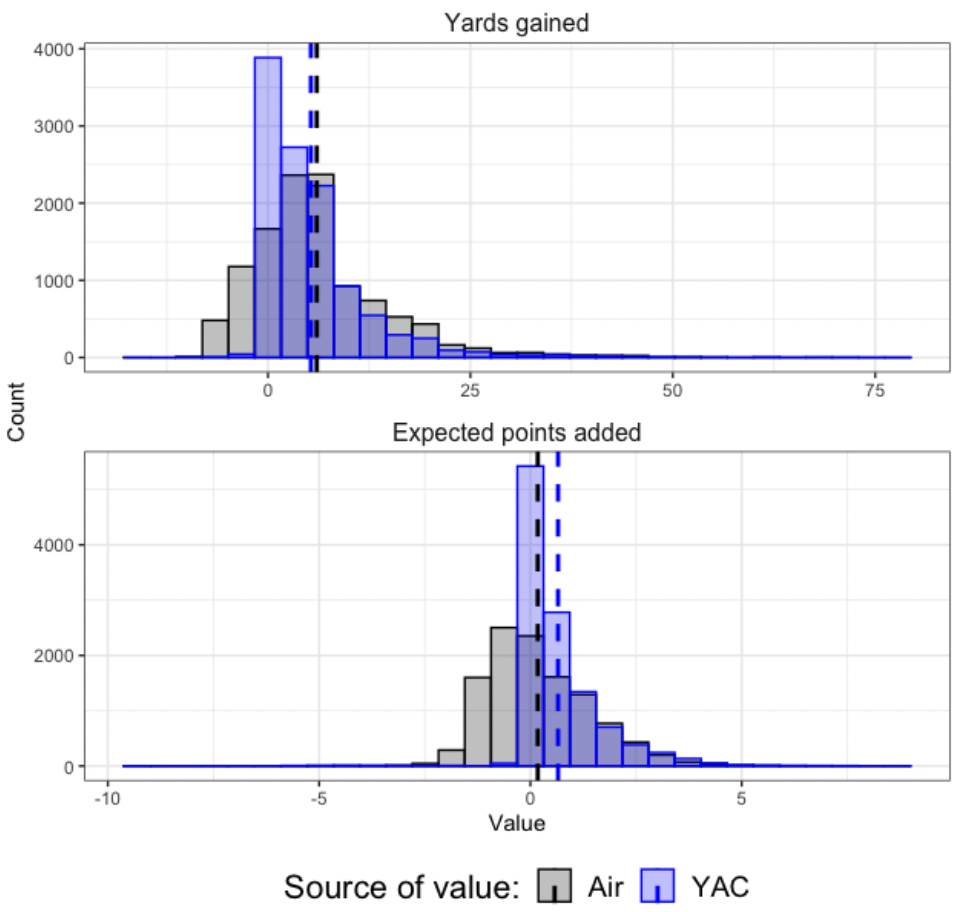}
    \caption{Comparison of distributions for yards gained during the 2018 NFL season through the air (in gray) versus yards after catch (in blue), along with the distribution for expected points added through the air versus after the catch obtained from the \texttt{nflfastR} \texttt{R} package \citep{carl2024nflfastr, r2024language}. Vertical dashed lines indicate the means for the respective distributions. Yards after catch represent a significant part of the value obtained through passing.}
    \label{fig:air-vs-yac}
\end{figure}

To illustrate our framework, we focus specifically on evaluating defensive pass coverage performance in the NFL.
In every passing play, there are two separate yardage components that contribute to the final spot where the ball is placed: (1) air yards, which represents the yards gained at the moment the ball is caught by the receiver, and (2) yards after catch (YAC), which represents the extra yards gained by the receiver after the catch is made. 
Figure \ref{fig:air-vs-yac} displays a comparison of the distributions of value added through the air versus YAC in terms of both yards gained and expected points added for receptions in the 2018 season. 
It is evident that YAC represents a significant portion of the value of passing plays and warrants investigation into the role defenders play in limiting YAC.

Using player tracking data provided by the NFL Big Data Bowl, we introduce a ghosting framework for analyzing the ability of defenders to limit YAC.
In particular, we compare the \amend{expected value of the observed} defensive player positioning and trajectory with a distribution of ghost defenders on passing plays.
Our approach leverages random forests for conditional density estimation (RFCDE) as the primary method and consists of two main components: (1) an estimate for the yards after catch distribution of the receiver, and (2) an estimate for the 2D location distribution of ghost defenders. For each model, we account for the \amend{multi}-dimensional features derived from player tracking data and play-level context. The benefits of our RFCDE approach are plentiful.
RFCDE allows for estimating the full distribution of our quantities of interest, rather than just a point prediction which differs from previous ghosting approaches.
Thus, we can quantify the uncertainty in the response for both univariate (yards after catch) and multivariate (2D defender location) cases, which is crucial for our framework.

\begin{figure}
    \centering
    \includegraphics[width = \textwidth]{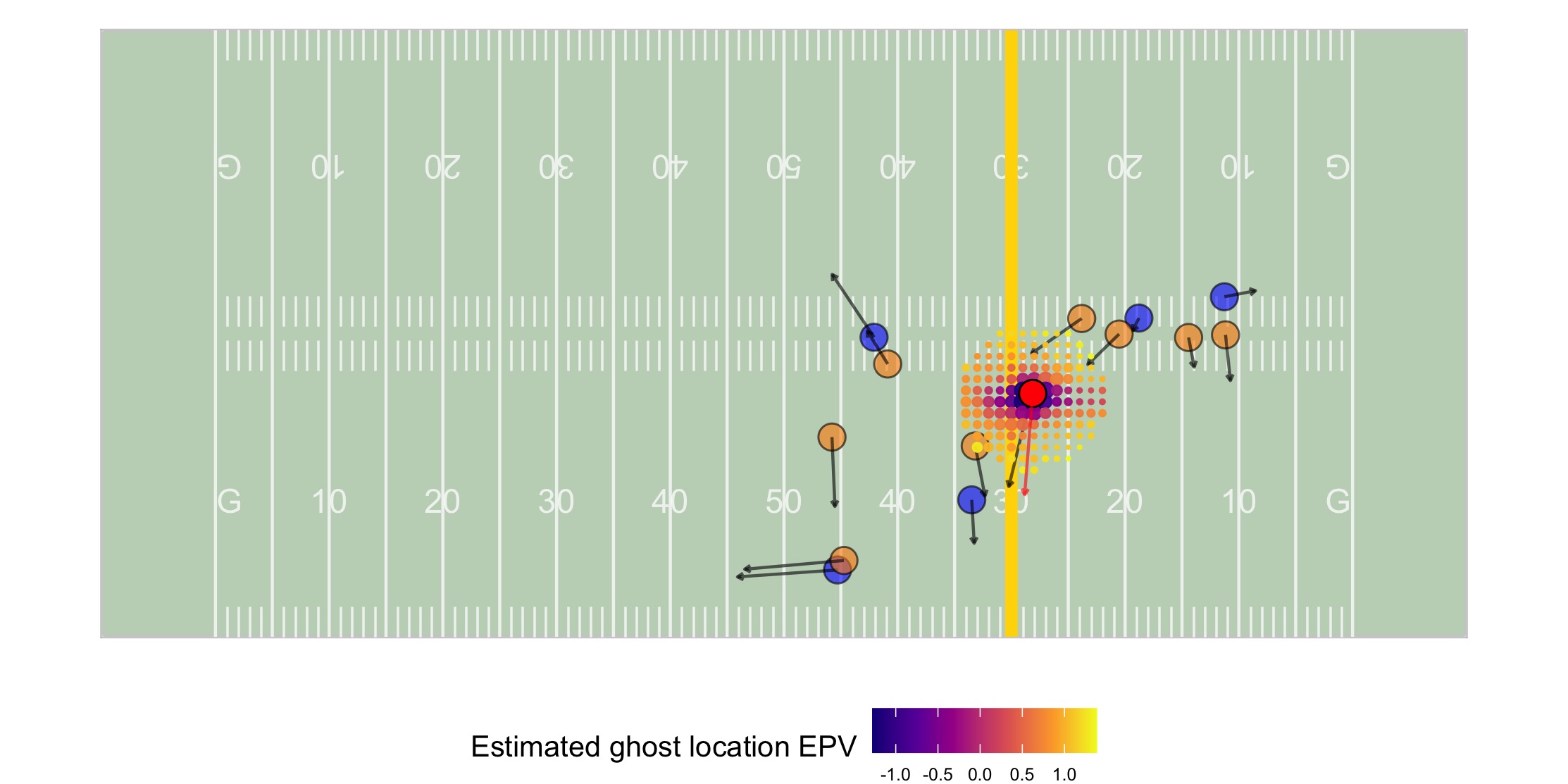}
    \caption{Example play heatmap of expected play values (EPV) over a grid of hypothetical ghost defender locations, with the observed nearest defender to the receiver at the moment of catch highlighted in red. The team on offense (in blue) is moving from right to left against the opposing defense (in orange). The first down line is highlighted in yellow, and the arrows indicates players' velocity vectors of movement. The size of the ghost grid points are proportional to the 2D RFCDE ghost defender location estimates, indicating which locations are more likely for the defender to be positioned at. The point color represents the associated location EPV with lower values (dark blue) corresponding to a better expected outcome for the defense, versus regions with higher positive EPV (yellow) that are better for the offense.}
    \label{fig:ex-play-epv-grid}
\end{figure}

Figure \ref{fig:ex-play-epv-grid} displays the culmination of our approach for an example play at the moment of catch, before the receiver was tackled by the nearest defender short of the first down line. The combination of estimates obtained from RFCDE allow us to observe expected play values across a grid of hypothetical ghost defender locations. Thus, our framework enables us to evaluate the positioning and trajectory of the defender in this play, relative to moving the player around the grid of ghost locations.  We note that even though our focus is on assessing defenders on passing plays, our framework can be extended to the within-play evaluation of players in other aspects of American football and other sports.

The rest of this paper is structured as follows.
In Section \ref{sec:data}, we provide an overview of the NFL tracking data. We describe our modeling approach and player evaluation framework in Section \ref{sec:methods}, and then demonstrate our framework with an example play along with a presentation of novel player and team metrics in Section \ref{sec:results}.
Finally, we conclude with a discussion of our findings in Section \ref{sec:discussion}.

\section{Data}
\label{sec:data}
For our analysis, we use the player tracking data supplied by the NFL Big Data Bowl 2021 \citep{howard2020nfl}.
The data are recorded at a frequency of 10 Hz (i.e. 10 times a second) and contain information for every passing play across all 17 weeks of the 2018 NFL regular season.
Of primary interest, the data provide x and y coordinates of players on the field at every frame within each play.
It is worth noting that offensive and defensive linemen information are omitted from the provided tracking data\amend{, resulting in fewer than 22 observed players on the field for each play}.
The data also record player attributes at each time step including speed, acceleration, distance traveled from previous frame, orientation, and angle of motion.
In addition, a vital feature in our data is the event label (e.g., ball snap, pass release, tackle, etc.) that corresponds to each moment of time within a play.

As an example, Table \ref{tab:tracking} presents the tracking data for \amend{a single player, Jaire Alexander, in} an example play (from Figure \ref{fig:ex-play-epv-grid}) during the 2018 season week 10 game between the Miami Dolphins and the Green Bay Packers.
For context, this play is a third down with six yards to go for the offense as Dolphins receiver DeVante Parker caught the forward pass thrown by his quarterback teammate, before being tackled by Packers cornerback Jaire Alexander. In Section \ref{sec:results}, we use this play as a case study to illustrate our proposed framework.

\begin{table}[t]
\caption{\amend{Example of tracking data for Green Bay Packers cornerback Jaire Alexander in a play during the Miami Dolphins versus Packers NFL game on November 11, 2018. The selected variables include frame identifier for each play (\texttt{frameId}); player location on the field (\texttt{x} and \texttt{y} coordinates); speed (\texttt{s}, in yards/second); acceleration (\texttt{a}, in yards/second$^2$); distance traveled from previous frame (\texttt{dis}, in yards); orientation (\texttt{o}, between 0 and 360 degrees); angle of motion (\texttt{dir}, between 0 and 360 degrees); and event label for each frame (\texttt{event}). The frames included are between the ball snap and when the tackle is made (the end-of-play event)}. \label{tab:tracking}}
\centering
\begin{tabular}{ccccccccl}
\hline
\texttt{frameId} & \texttt{x} & \texttt{y} & \texttt{s} & \texttt{a} & \texttt{dis} & \texttt{o} & \texttt{dir} & \texttt{event} \\ 
\hline
11 & 84.63 & 43.12 & 0.00 & 0.00 & 0.00 & 92.26 & 238.61 & ball\_snap \\ 
12 & 84.63 & 43.13 & 0.00 & 0.00 & 0.00 & 92.26 & 250.10 & None \\ 
\vdots & \vdots & \vdots & \vdots & \vdots & \vdots & \vdots & \vdots & \vdots \\
38 & 81.45 & 33.11 & 8.05 & 1.94 & 0.81 & 167.45 & 180.08 & pass\_forward \\ 
39 & 81.46 & 32.28 & 8.29 & 1.90 & 0.84 & 162.66 & 178.47 & None \\ 
\vdots & \vdots & \vdots & \vdots & \vdots & \vdots & \vdots & \vdots & \vdots \\
43 & 81.70 & 28.79 & 9.01 & 0.90 & 0.90 & 167.24 & 174.88 & pass\_arrived \\ 
44 & 81.78 & 27.88 & 9.13 & 0.49 & 0.91 & 167.24 & 174.72 & None \\ 
\vdots & \vdots & \vdots & \vdots & \vdots & \vdots & \vdots & \vdots & \vdots \\
51 & 81.87 & 21.40 & 8.93 & 3.03 & 0.90 & 198.63 & 184.53 & pass\_outcome\_caught \\ 
52 & 81.78 & 20.52 & 8.75 & 3.34 & 0.89 & 205.36 & 186.89 & None \\ 
\vdots & \vdots & \vdots & \vdots & \vdots & \vdots & \vdots & \vdots & \vdots \\
55 & 81.30 & 18.03 & 8.05 & 3.66 & 0.82 & 205.80 & 193.49 & tackle \\ 
\hline
\end{tabular}
\end{table}


In total, there are 19,239 passing plays where the play outcome includes complete pass, incomplete pass, intercepted pass, or quarterback sack. Since the primary focus of this paper is the assessment of defender ability to limit yards after catch, we only consider plays that result in a complete pass \amend{thrown by a quarterback} outside of the end zone (i.e., the receiver catches the forward pass by the quarterback with yards remaining to their target end zone). We take a similar approach to \cite{yurko2020going} to construct tracking data features \amend{in Table \ref{tab:player-vars} that capture information different types of players involved at relevant events within a play. In particular, we split the players into four groups: quarterback (\texttt{qb}), receiver (\texttt{rec}), offense excluding quarterback (\texttt{offX}), and defense (\texttt{defX}), where we order the offensive and defensive players based their Euclidean distance to the receiver at the moment of catch (e.g., {\tt def1} denotes the \textit{closest} defender).  All of the features are computed at the moment of catch, except for information about the quarterback which is measured at the moment the ball is thrown. After pre-processing our data to only consider relevant passing plays without missing values for the considered features in Table \ref{tab:player-vars}, our analysis is limited to 10,363 completed passes. We have no reason to suspect the data are not missing at random.}

\begin{table}[!ht]
\centering
\caption{List of features constructed from player tracking data \amend{for the four different player groups:  quarterback (\texttt{qb}), receiver (\texttt{rec}), offense excluding quarterback (\texttt{offX}), and defense (\texttt{defX})}. We order the offensive and defensive players based their Euclidean distance to the receiver at the moment of catch (e.g., {\tt def1\_x\_adj} denotes how many yards away the \textit{closest} defender is from the receiver's target endzone). Since there are multiple players involved in a play, there were can be multiple versions of the same variable computed (e.g., \texttt{qb\_s}, \texttt{rec\_s}, \texttt{def1\_s}, etc.)}
\label{tab:player-vars}
\centering
    \begin{tabular}{p{6cm} p{7cm}}
        \hline
        Features (relevant positions) & Description  \\
        \hline
        \hline
        \texttt{x\_adj} (\texttt{rec}, \texttt{off}, \texttt{def}) & Horizontal yards from the receiver's target endzone. \\
        \hline
        \texttt{y\_adj} (\texttt{rec}, \texttt{off}, \texttt{def}) & Vertical yards from the center of field with respect to the target endzone, where positive values indicate left side while negative values indicate right side.     \\
        \hline
        \texttt{dir\_endzone} (\texttt{rec}, \texttt{off}, \texttt{def}) & Absolute value of the direction a player is moving with respect to the target endzone, where 0 indicates that the player is moving towards the endzone, while positive degrees between 0 and 180 indicate that the player is moving to the left or right. \\
        \hline
        \texttt{o\_endzone} (\texttt{rec}, \texttt{off}, \texttt{def}) & Absolute value of the orientation a player's shoulder pads are facing with respect to the target endzone, where 0 indicates that the player is facing the endzone while positive degrees between 0 and 180 indicate that the player is facing left or right. \\
        \hline
        \texttt{x\_adj\_from\_first\_down} (\texttt{rec}) & Distance (in yards) the receiver is from the first down line (or endzone in goal to go downs) where positive values denote the yards to go while negative values indicate yards past the first down line. \\
        \hline
        \texttt{s} (\texttt{qb}, \texttt{rec}, \texttt{off}, \texttt{def}) & The speed (in yards/second) a player is moving at. \\
        \hline
        \texttt{x\_adj\_change} (\texttt{qb}, \texttt{off}, \texttt{def}) & Horizontal displacement between a player and the receiver according to \texttt{x\_adj} values. \\
        \hline
        \texttt{y\_adj\_change} (\texttt{qb}, \texttt{off}, \texttt{def}) & Absolute value of vertical displacement between a player and the receiver according to \texttt{y\_adj} values. \\
        \hline
        \texttt{dist\_to\_rec} (\texttt{qb}, \texttt{off}, \texttt{def}) & Euclidean distance between a player and the receiver. \\
        \hline
        \texttt{dir\_wrt\_rec\_diff} (\texttt{off}, \texttt{def}) & Minimal absolute difference between a player's direction of movement and the angle between the player and the receiver. \\
        \hline
        \texttt{o\_wrt\_rec\_diff} (\texttt{off}, \texttt{def}) & Minimal absolute difference between a player's orientation and the angle between the player and the receiver. \\
        \hline
    \end{tabular}
\end{table}

\section{Methods}
\label{sec:methods}

Our framework for evaluating the positioning of defenders at the moment of catch relies on distribution estimates for (1) the yards after catch gained by a receiver and (2) the 2D location of ghost defenders. \amend{Instead of relying on parametric assumptions for these distributions, we use random forests for conditional density estimation \citep[RFCDE;][]{dalmasso2020conditional, pospisil2018rfcde}}. RFCDE is a flexible nonparametric tree-based approach that can handle the \amend{multi}-dimensional tracking data features described in Table \ref{tab:player-vars}. Here, the features are accounted for via weighted kernel density estimation, where the weights are determined by the terminal nodes of the trees in a random forest \citep{breiman2001random}. 

Following the notation in \cite{pospisil2019f}, let $\theta_t$ represent the tree structure for tree $t$ and $R(X^*, \theta_t)$ denote the feature space region covered by its terminal leaf node for observation $X^*$. For observation $X^*$, the weight for each observation $i = 1, \dots, n$ in the training data across $T$ trees is calculated as
\begin{equation}
\label{eq:weights}
    w_i(X^*) = \frac{1}{T} \sum_{t = 1}^T \frac{\mathbbm{1}(X_i \in R(X^*, \theta_t))}{\sum_{i=1}^n \mathbbm{1}(X_i \in R(X^*, \theta_t))}.
\end{equation}
Unlike traditional random forests for regression or classification, the tree splits in RFCDE are determined to minimize a loss specific to conditional density estimation \citep{izbicki2017converting}. \amend{
Ultimately, the weights in Equation \ref{eq:weights} are used to perform weighted kernel density estimation},
\begin{equation}
    \hat{f}(Y \mid X^*) = \frac{1}{\sum_{i=1}^n w_i(X^*)} \sum_{i=1}^n w_i(X^*) K_h(Y_i - Y),
\end{equation}
where $K_h$ is a kernel function (e.g., Gaussian), with bandwidth $h$ chosen using plug-in methods. The RFCDE estimate $\hat{f}(Y \mid X^*)$ determines how ``close'' each of the training data observations are to the point of interest based on whether they belong to the same leaf node in the collection of $T$ trees. This provides a flexible way to capture \amend{multi}-dimensional features and interactions for the purpose of conditional density estimation. RFCDE has been successfully demonstrated before in the context of modeling yards gained by NFL running backs using tracking data \citep{yurko2020going}. \amend{We refer the reader to  \cite{dalmasso2020conditional} for a complete overview and comparison of RFCDE with other CDE approaches. Additionally, as discussed by \cite{dalmasso2020conditional} and \cite{pospisil2018rfcde}, RFCDE can scale for multivariate responses.} For the remainder of this section, we rely on the RFCDE software described in detail by \cite{dalmasso2020conditional} and use $T = 500$ trees to estimate our distributions of interest.

\subsection{Expected play value at the moment of catch}
\label{sec:epv-yac}

Our quantity of interest is the expected play value (EPV) for the offensive team at the moment of catch. Let $V$ be some measure of play-value, such as the output of the previously mentioned expected points (EP) or win probability (WP) model described in \cite{yurko2019nflwar}. Then our quantity of interest is
\begin{equation}
\label{eq:epv}
    EPV_{catch} = \mathbb{E}[V \mid X_{catch}],
\end{equation}
where $X_{catch}$ contains covariate information observed at (or up to) the moment of time the receiver makes the catch, such as the features described in Table \ref{tab:player-vars}.  

In practice, the play value $V$ is the result of a complex utility function $g()$ of the receiver's ending yard line $Y$. A receiver's ending yard line is simply a combination of their field position at the moment of catch and their yards gained after the catch (YAC). In other words, modeling the receiver's ending yard line is equivalent to modeling the receiver's YAC conditioned on their known starting position at the moment of catch. In order to compute $EPV_{catch}$, we need to integrate over the conditional density $f$ for the receiver's ending yard line $Y$ given covariate information $X_{catch}$,

\begin{equation}
    \label{eq:epv-integral}
    EPV_{catch} = \int g(Y)  f(Y \mid X_{catch}) dY.
\end{equation}

As for measuring the utility $g()$, we use the expected points approach of \cite{yurko2019nflwar}, which relies on a multinomial logistic regression model to predict that next scoring event given a play's context (e.g., down, yards to go, yardline, etc.). This is just one example of an EP model, with a variety of approaches in the literature each possessing their own problems and limitations (see \cite{brill2025analytics} for a description). However, the choice of the EP approach does not matter in the context of our framework. As long as the choice of the utility function $g()$ is a function of the ending yard line, then our considered EP model can be replaced but our approach will remain valid.

To compute $EPV_{catch}$ in Equation $\ref{eq:epv-integral}$, we need an estimate for the conditional density of the receiver's ending yard line, $\hat{f}(Y \mid X_{catch})$. To this end, we use RFCDE to estimate this density as a function of \amend{multi}-dimensional tracking data features $X_{catch}$. Although the trees within RFCDE implicitly perform variable selection, we use leave-one-week-out cross validation (LOWO CV) to assess the out-of-sample performance of the RFCDE estimate $\hat{f}(Y \mid X_{catch})$ for a varying set of features based on the included number of defensive and offensive players. \amend{This is necessary due to the inherent difficulty in estimating a CDE loss compared to traditional tree-splitting features (e.g., mean squared error), which may lead to poor tree splits by RFCDE in the presence of many noisy features \citep{dalmasso2020conditional}.} We consider three evaluation metrics for the YAC model: (1) CDE loss considered in \cite{dalmasso2020conditional}, (2) root mean squared error (RMSE) between the observed YAC with RFCDE mean, and (3) RMSE with RFCDE mode. We consider both the mean and mode relevant since the YAC distributions will likely be skewed and exhibit non-symmetrical behavior. We observe that the YAC density estimate's LOWO CV performance do not improve after accounting for information about the nearest defender (\texttt{def1}), in addition to accounting for the receiver and quarterback information in Table \ref{tab:player-vars} (see Figure S1 in Supplementary Material). This simplicity in the considered number of defenders is likely driven by the fact we are only modeling information at the moment of catch (as well as quarterback information at the moment when the pass is released), with challenges for extending this approach discussed later in Section \ref{sec:discussion}.

\amend{To provide additional context on the performance of RFCDE in this setting, we perform a sensitivity analysis to observe the performance of RFCDE as a function of the amount of training data. We evaluate the YAC RFCDE performance using the three aforementioned metrics by incrementally increasing the amount of training data by one week's worth of plays (i.e., start with week 1, then increase to weeks 1--2, weeks 1--3, and so on) while using the final week 17 as the test data. We observe that the performance of the YAC RFCDE stabilizes after about eight weeks of data (see Figure S3 in Supplementary Material).}

For each of our considered 10,363 passing plays, we obtain $\widehat{EPV}_{catch}$ using RFCDE for yards after catch trained on all plays with the quarterback (\texttt{qb}), receiver (\texttt{rec}), and closest defender (\texttt{def1}) features listed in Table \ref{tab:player-vars}\amend{, resulting in twenty features}. Specifically, for each play, we generate conditional density estimates for the receiver's YAC (or ending yard line) $\hat{f}(Y \mid X_{catch})$ over a grid of possible values in increments of one yard that range from -10 yards gained (determined based on the yardage distribution displayed in Figure \ref{fig:air-vs-yac}) to the maximum possible yards gained given the location of the catch (i.e., distance to the target end zone). Similar to the RFCDE approach in \cite{yurko2020going}, we include a ``padding'' of additional two yards to the maximum possible value for a better estimate of the receiver reaching the target end zone. Since we are estimating the conditional density in a discrete-like manner, for a given play, we normalize the conditional density estimates $\hat{f}(Y \mid X_{catch})$ over the considered grid of values so that they add up to one. Our estimate for $\widehat{EPV}_{catch}$ is then based on the summation of these discrete probability estimates multiplied by their respective utility function $g(Y)$ values.

\subsection{\amend{Framework for estimating defender value at catch}}
\label{sec:epv-ghost}

While the quantity in Equation \ref{eq:epv-integral} gives us the receiver's expected value at the moment of catch, our goal is to assess the spatial positioning of defenders in limiting the receiver's value added after the catch. This amounts to the difference between $EPV_{catch}$ and the expected value if we replaced the defender of interest with a hypothetical ghost defender, which we denote as $EPV_{catch}^{ghost}$. With regards to Equation \ref{eq:epv-integral}, this means replacing the observed covariate information $X_{catch}$ with a ghost version $\tilde{X}_{catch}$ \amend{such that

\begin{equation}
    \label{eq:ghost-epv-integral}
    EPV_{catch}^{ghost} = \int g(Y)  f(Y \mid \tilde{X}_{catch}) dY,
\end{equation}
where we rely on our conditional density estimate for the ending yard line described in the previous section for $f$ that takes in the ghost features represented by $\tilde{X}_{catch}$.

In order to evaluate the defender positioning and trajectory at the moment of catch for a given play, we simply take the difference between the player's and ghost's EPV values, which we represent with $\delta_{catch}$, where
\begin{equation}
    \label{eq:epv-change}
    \delta_{catch} = EPV_{catch} - EPV_{catch}^{ghost}.
\end{equation}

However, we recognize that there is a distribution of hypothetical ghost defenders for which we want to compare the actual player against. Thus, we are interested in estimating the expected value of $\delta_{catch}$, which involves integrating over the possible ghosts $\mathcal{G}$ with access to the conditional density $h$ of the ghosts given covariate information $X_{catch}^{-\texttt{def1}}$ that excludes the features on the nearest defender to the receiver. That is,

\begin{equation}
    \label{eq:exp-epv-change}
    \mathbb{E}[\delta_{catch}] = \int_{\gamma \in \mathcal{G}} (EPV_{catch} - EPV_{catch}^{\gamma} )h(\gamma \mid X_{catch}^{-\texttt{def1}}) d \gamma.
\end{equation}

We treat the distribution of ghost defenders as five-dimensional, decomposed into two parts: (1) 2D defender location on the field; and (2) 3D defender trajectory consisting of their speed, direction of movement, and shoulder orientation. For ease of notation, we let $\ell$ represent the location tuple ($\texttt{x}, \texttt{y}$) across the grid of possible locations $\mathcal{L}$ and $v$ represent the 3D trajectory vector ($\texttt{s}, \texttt{dir}, \texttt{o}$) from the possible defender trajectories $\mathcal{V}$. Rather than attempting to estimating the joint 5D density of locations and trajectories, we rewrite Equation \ref{eq:exp-epv-change} by conditioning on the location $\ell$ prior to integrating over the trajectory vectors,
\begin{equation}
    \label{eq:exp-full-epv-change}
    \mathbb{E}[\delta_{catch}] = \int_{\ell \in \mathcal{L}} \Big( \int_{v \in \mathcal{V}} (EPV_{catch} - EPV_{catch}^{\ell, v} ) p(v \mid \ell) dv \Big) h(\ell \mid X_{catch}^{-\texttt{def1}}) d \ell.
\end{equation}
In order to estimate $\mathbb{E}[\delta_{catch}]$ in Equation \ref{eq:exp-full-epv-change}, we need two additional conditional density estimates: (1) 2D defender location $h(\ell \mid X_{catch}^{-\texttt{def1}})$, and (2) 3D defender trajectory $p(v \mid \ell)$.

For the 2D nearest defender location conditional density estimate $h$, we use a 2D RFCDE that relies on the same multi-dimensional tracking data features as before but with the information about the nearest defender removed (i.e., $X_{catch}^{-\texttt{def1}}$ is a modified version of $X_{catch}$ but without the \texttt{def1} features.) This results in a vector of ten features. We also explore the performance of the 2D RFCDE with additional features that account for the receiver's nearest teammate (\texttt{off1}) and the second closest defender (\texttt{def2}). However, based on the LOWO CV results, we observe that only relying on information about the receiver and quarterback is optimal as measured by cross-entropy and the Euclidean distance between the observed defender location with the 2D RFCDE mode (see Figure S2 in the Supplementary Material). We also conduct the same type of sensitivity analysis for the 2D RFCDE as we perform for the YAC RFCDE by gradually increasing the number of weeks in the training data. Unsurprisingly, the 2D RFCDE appears to require more weeks to stabilize in performance with similar values observed after twelve weeks of data (see Figure S4 in Supplementary Material). We recognize that there is potentially other information that is relevant in modeling a defender's location, but leave expanding on additional features with larger datasets than our considered sample for future work. 

Hypothetically, one could use a 3D RFCDE to estimate the trajectory density $p$. Due to the inherent difficulty of this task, we instead use a sampling strategy similar to \cite{deshpande2020expected} by resampling the observed trajectory for the nearest defenders at the moment of catch. Rather than naively resampling without conditioning on any information about the defender's location, we weight defender trajectories from the observed dataset based on similar distances with the receiver as the considered 2D location $\ell$. Specifically, we compute the sampling weights for observation $i$'s 3D trajectory (speed, direction, and orientation) for location $\ell$ as
\begin{equation}
\label{eq:sample-weights}
    \omega_{i,\ell} = \frac{1}{|\texttt{def1\_dist\_to\_rec}_{i} - \texttt{def1\_dist\_to\_rec}_{\ell}|},
\end{equation}
i.e., observations with similar distances to the receiver as the hypothetical location $\ell$ will have a higher probability of being sampled. We consider this sampling approach as a reasonable starting point for preserving the correlation structure between the 3D trajectory vector of defenders. We leave the exploration of properly modeling the trajectory information for future work that one could develop to plug-in to our modular framework. 

Our full process for estimating $\mathbb{E}[\delta_{catch}]$ in Equation \ref{eq:exp-full-epv-change} proceeds as follows:
\begin{enumerate}
    \item For each 2D ghost location $\ell$, resample $B$ trajectory vectors using the weightes $\omega_{i, \ell}$ in Equation \ref{eq:sample-weights}.
    \item Update the nearest defender features with the ghost location $\ell$ and sampled trajectory $v_b$ to create $\tilde{X}^{\ell, b}_{catch}$ and compute $EPV_{catch}^{\ell, b}$ following Equation \ref{eq:ghost-epv-integral}.
    \item Integrate over the $B$ resampled defender trajectories and 2D ghost location RFCDE to estimate $\mathbb{E}[\delta_{catch}]$ with,
    \begin{equation}
        \int_{\ell \in \mathcal{L}} \Big( \frac{1}{B} \sum_{b}^B (EPV_{catch} - EPV_{catch}^{\ell, b} ) \Big) h(\ell \mid X_{catch}^{-\texttt{def1}}) d \ell.
    \end{equation}
\end{enumerate}
Additionally, we provide a comparison of the Equation \ref{eq:exp-full-epv-change} estimates with our weighted sampling strategy versus using the defender's observed trajectory (i.e., ignoring the trajectory dependence on the change to location $\ell$) in Figure S5 of the Supplementary Material.

In addition to the estimate for $\mathbb{E}[\delta_{catch}]$, our novel framework results in a multitude of quantities such as a distribution of $EPV_{catch}^{\ell, b}$ values to compare with the $EPV_{catch}$ estimate from the observed defender's location and trajectory at the moment of catch. Ultimately, our estimates for $\mathbb{E}[\delta_{catch}]$ measure the value of the defender's positioning at the catch relative to a distribution of hypothetical defenders.} We are able to aggregate these values for players and teams across the observed season of data to create new valuation metrics based on player tracking data. \amend{While our framework may seem computationally burdensome, after fitting the necessary RFCDE quantities, it is a parallel computing problem that took about 36 hours to generate all of the desired quantities across the entire of season of passing plays (distributed across 12 cores on a 2019 iMac).} All of the code to implement our framework are available on GitHub at \url{https://github.com/ryurko/nfl-ghosts}.

\section{Results}
\label{sec:results}
\subsection{Example play valuations}
\label{sec:ex-play-results}

\begin{figure}
    \centering
    \includegraphics[width = \textwidth]{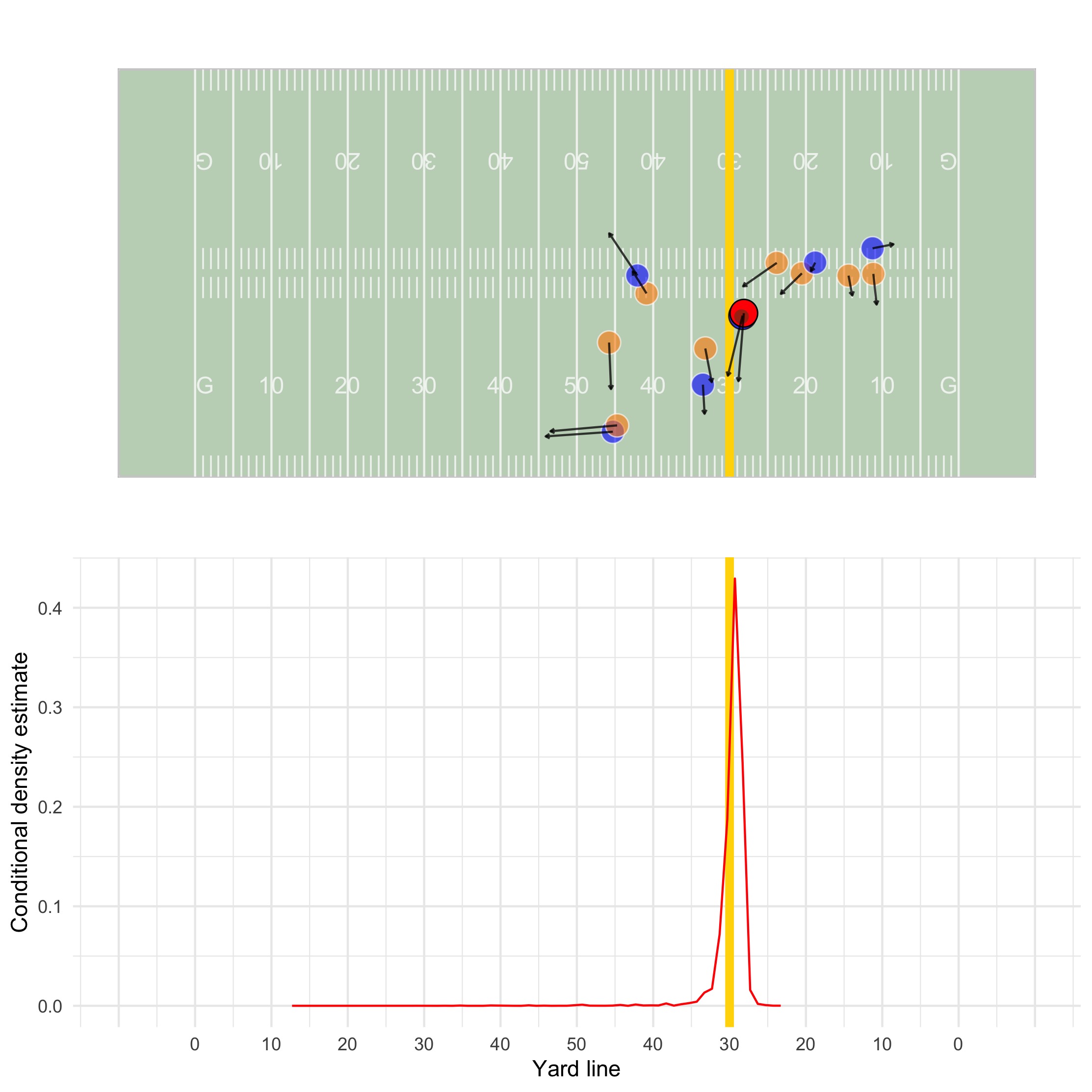}
    \caption{Visualization of the example play featuring nearest defender Jaire Alexander (in red), offensive players (in blue), and other defensive players (in orange) at the moment of catch, along with the corresponding RFCDE for yards after catch. The first down line (in gold) is displayed for reference, indicating that \amend{most of} the mass of the conditional density estimate for YAC is before the first down line (as the play direction is from right to left).}
    \label{fig:ex-play-yac-rfcde}
\end{figure}

We demonstrate our framework with the example play mentioned in Section \ref{sec:data}, where Green Bay Packers cornerback Jaire Alexander tackled the receiver short of the first down line. First, we estimate the expected value at the moment of catch ($\widehat{EPV}_{catch}$) using the RFCDE for yards after catch described in Section \ref{sec:epv-yac}. Figure \ref{fig:ex-play-yac-rfcde} displays the locations of players on the field for the example play along with the CDE of the receiver's YAC based on the observed tracking data at the moment of catch. We observe that \amend{most of} the mass of the YAC density estimate is behind the first down line, which is beneficial to the defense. Using this CDE along with the previously described EP model, \amend{we estimate $\widehat{EPV}_{catch} = -1.35$} as the expected play value when the receiver catches the ball. Here, notice that the expected value for the offense when the receiver catches the ball based on the observed player tracking data, with the considered features that include the nearest defender, is negative. In other words, the defense is more likely to score next based on information observed up to the moment of the ball being caught. Ultimately, this play does result in the nearest defender tackling the receiver short of the first down, which leads to the offensive team turning the ball over to the defense with a punt.

\begin{figure}
    \centering
    \includegraphics[width = \textwidth]{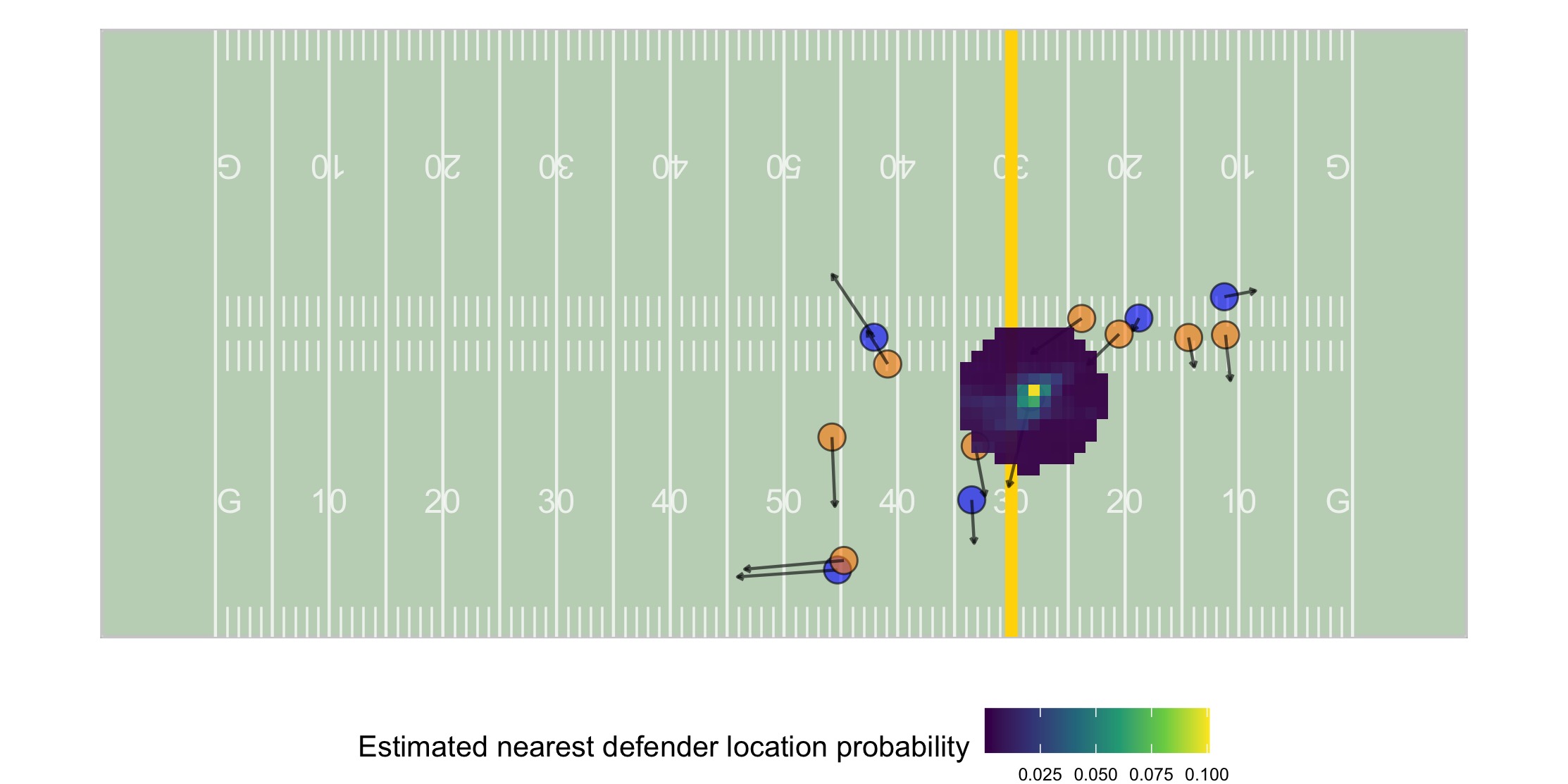}
    \caption{Heatmap display of 2D RFCDE for nearest defender positioning in the example play (with the actual nearest defender removed). \amend{The CDE estimates across the grid are normalized to sum to one, yielding discrete probability estimates ranging from low (dark blue) to high (yellow).}}    \label{fig:ex-play-ghost-grid}
\end{figure}

Next, we use our proposed framework in Section \ref{sec:epv-ghost} to evaluate the positioning and trajectory of the nearest defender to the receiver relative to the ghost distribution baseline. Figure \ref{fig:ex-play-ghost-grid} displays the 2D RFCDE output for the example play \amend{(displayed on a normalized probability scale), showing which locations on the field are more likely for the closest defender to be positioned. We observe the most likely location (in yellow) is located extremely close to the receiver, and also notice} an asymmetric distribution for the ghost defender locations, supporting the need for a flexible technique to model player locations. 

\amend{
Following our framework Section \ref{sec:epv-ghost}, we generate a distribution of hypothetical ghost EPV values across possible defender locations with $B = 100$ resampled trajectory vectors. Figure \ref{fig:ghost-epv-distr} displays the distribution of estimated $EPV_{catch}^{\ell, b}$ values in the example play compared to $\widehat{EPV}_{catch} = -1.35$ for Jaire Alexander's observed positioning and trajectory at the time of catch. We observe a multi-modal distribution of ghost values with a large mass below 0 (better for the defense) and another mass above 0 (better for the offense). Although we do not present distributions for all plays, we observed similar behavior when spot-checking examples which we interpret as indication of modes showing better or worse defensive positioning. In evaluating Alexander's positioning, it is clear that his defensive positioning and trajectory are a top percentile performance for a defender at this moment in time. While our results are motivated by the quantity in Equation \ref{eq:exp-full-epv-change}, our framework does allow us to create percentile-based measures that could be used to evaluate the player's positioning relative to a distribution of ghosts. 

\begin{figure}
    \centering
    \includegraphics[width = \textwidth]{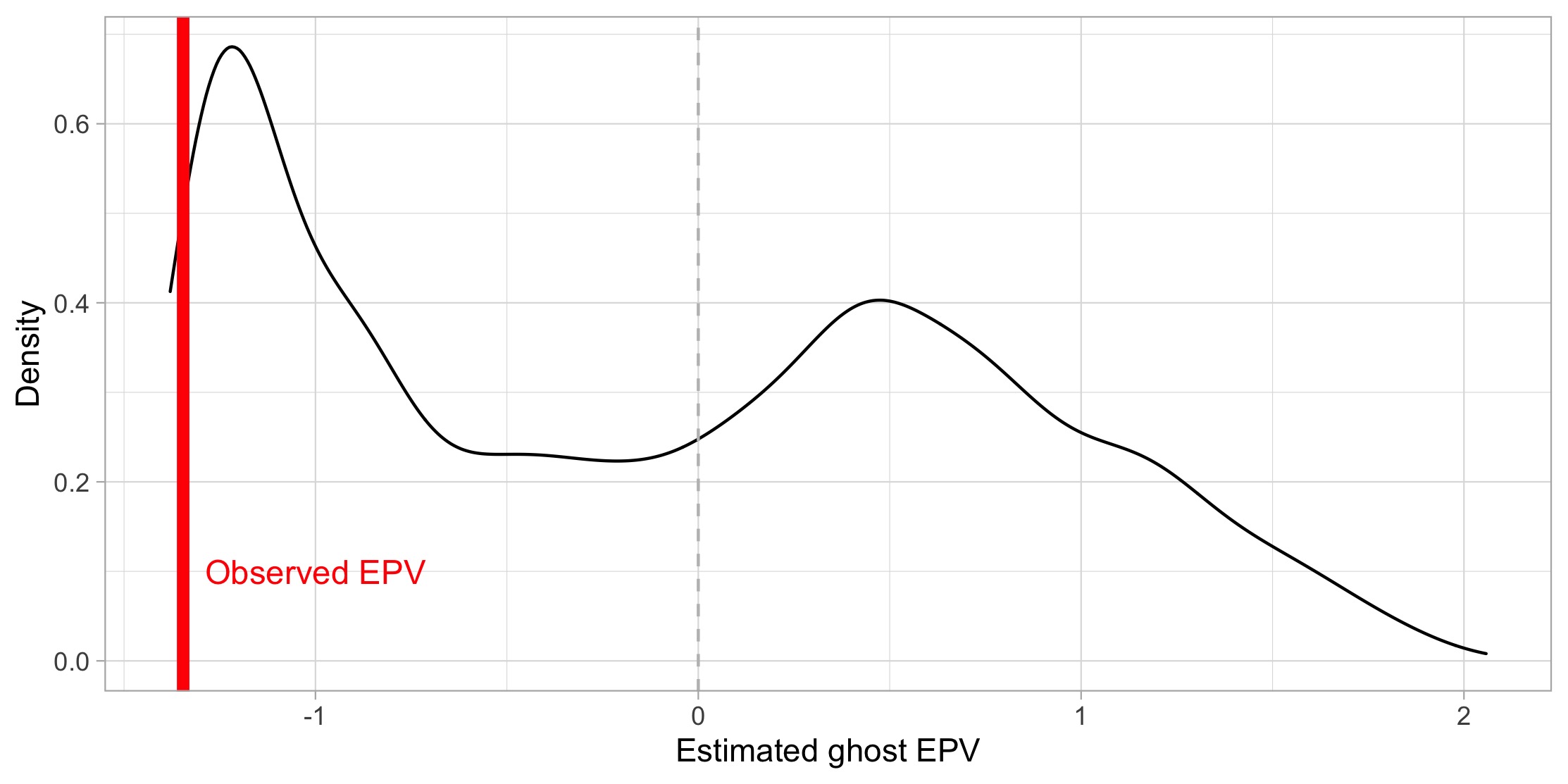}
    \caption{\amend{Distribution of estimated $EPV_{catch}^{\ell, b}$ values in the example play compared to $\widehat{EPV}_{catch} = -1.35$ (denoted by the red line) for Jaire Alexander's observed positioning and trajectory at the time of catch. We observe a multi-modal distribution of ghost values with a large mass below 0 (gray dashed line) indicating better defensive value, and another mode above 0 indicating better offensive value.}}    
    \label{fig:ghost-epv-distr}
\end{figure}

Figure \ref{fig:ex-play-epv-grid} displays the estimates for  $EPV_{catch}^{\ell}$, which are based on averaging across the $B$ sampled trajectory vectors for each grid location. Here, the locations with the lowest $\widehat{EPV}_{catch}^{\ell}$ values, which are more beneficial to the defense, are located right on top of the receiver, which is where the nearest defender is actually observed. We also see that the locations further away from the actual location of the defender are more beneficial for the offense. Interestingly, we observe asymmetry in the locations with locations that appear more orthogonal to the receiver's direction of movement are more optimal relative to trailing or directly going towards the receiver. This may suggest differences in tackling strategy, but the black-box nature of our RFCDEs leads more to be desired from this perspective that we leave for future study. 

As evidenced by both Figures \ref{fig:ex-play-epv-grid} and \ref{fig:ghost-epv-distr}, the defender is positioned in an optimal location at the moment of catch. This is reinforced in Figure \ref{fig:ex-play-epv-diff-grid} which displays a heatmap of the estimated $\delta_{catch}^{\ell}$ values showing the difference in value between the defenders positioning with the hypothetical ghost locations (averaged across the $B$ sample trajectory vectors). Finally, we integrate over the 2D location RFCDE and obtain $\mathbb{E}[\delta_{catch}] \approx -1.22$ as the estimate for this example play. This indicates that the observed nearest defender's spatial positioning and trajectory is worth approximately over one point in favor of the defense, relative to the ghost defender distribution. Ultimately, our framework provides us with the first known point-based evaluation system of a player's spatial positioning and trajectory from tracking data. 

\begin{figure}
    \centering
    \includegraphics[width = \textwidth]{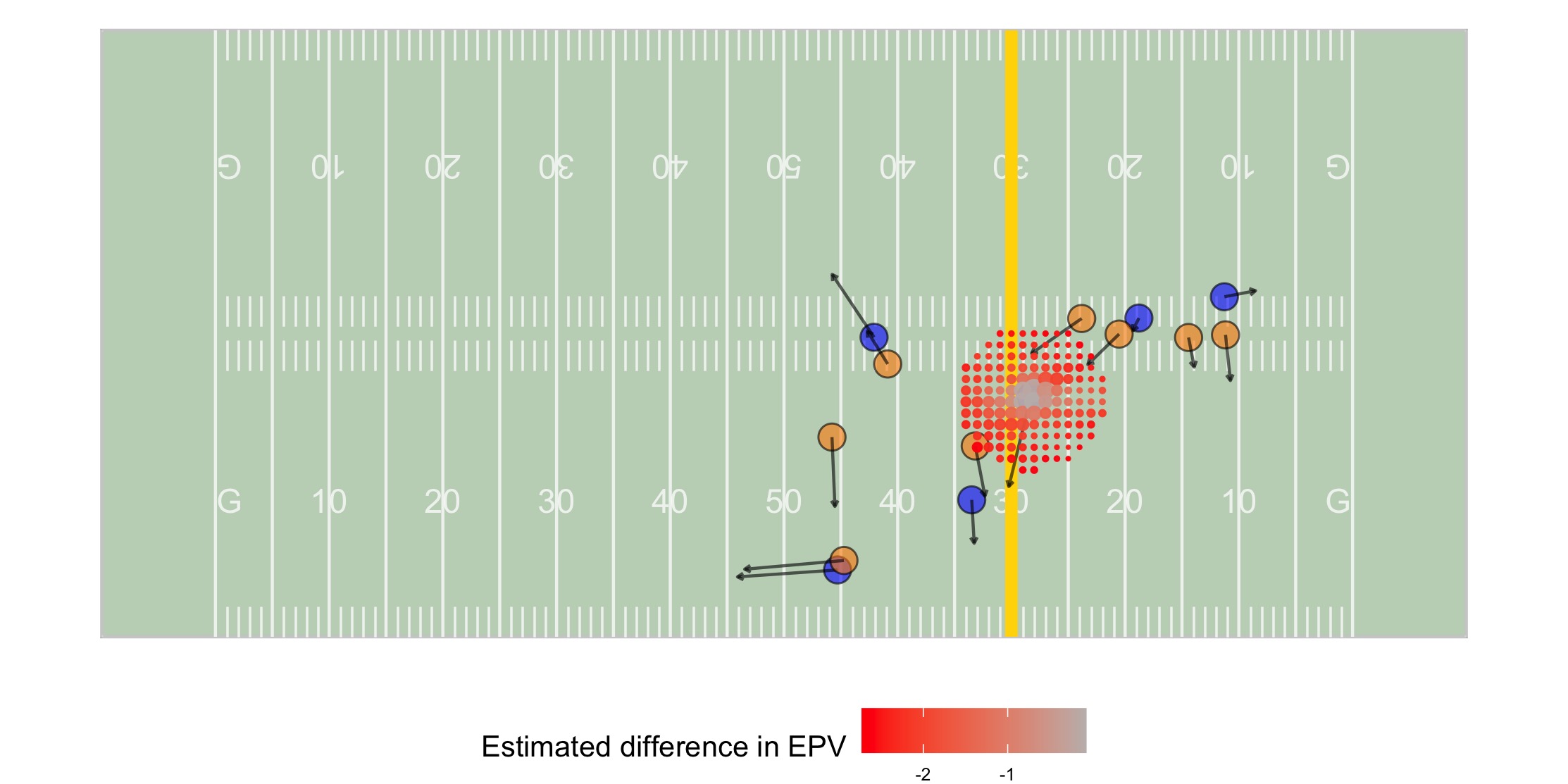}
    \caption{\amend{Example play heatmap of estimated $\delta_{catch}^{\ell}$ values indicating the difference in value between the defenders positioning with the hypothetical ghost locations (averaged across the $B$ sample trajectory vectors). Negative (red) values indicate locations that were worse than Alexander's from the defensive point of view. Alexander is removed from the visual, but the values closer to 0 (gray) are indicative of where he was located at the time of catch. The size of the ghost grid points are proportional to the 2D RFCDE ghost defender location estimates, indicating which locations are more likely for the defender to be positioned at.}}
    \label{fig:ex-play-epv-diff-grid}
\end{figure}

}

\subsection{Player and team performance}
\label{sec:player-team-results}

For each of the 10,363 considered plays in the data, we apply the same procedure as outlined in Section \ref{sec:ex-play-results} \amend{to compute the $\mathbb{E}[\delta_{catch}]$ values defined in Equation \ref{eq:exp-full-epv-change}}. This allows us to assess the positioning and trajectory of the closest defender to the receiver at the moment of catch relative to a baseline ghost defender over the entire 2018 NFL regular season.

\begin{figure}
    \centering
    \includegraphics[width = 0.6\textwidth]{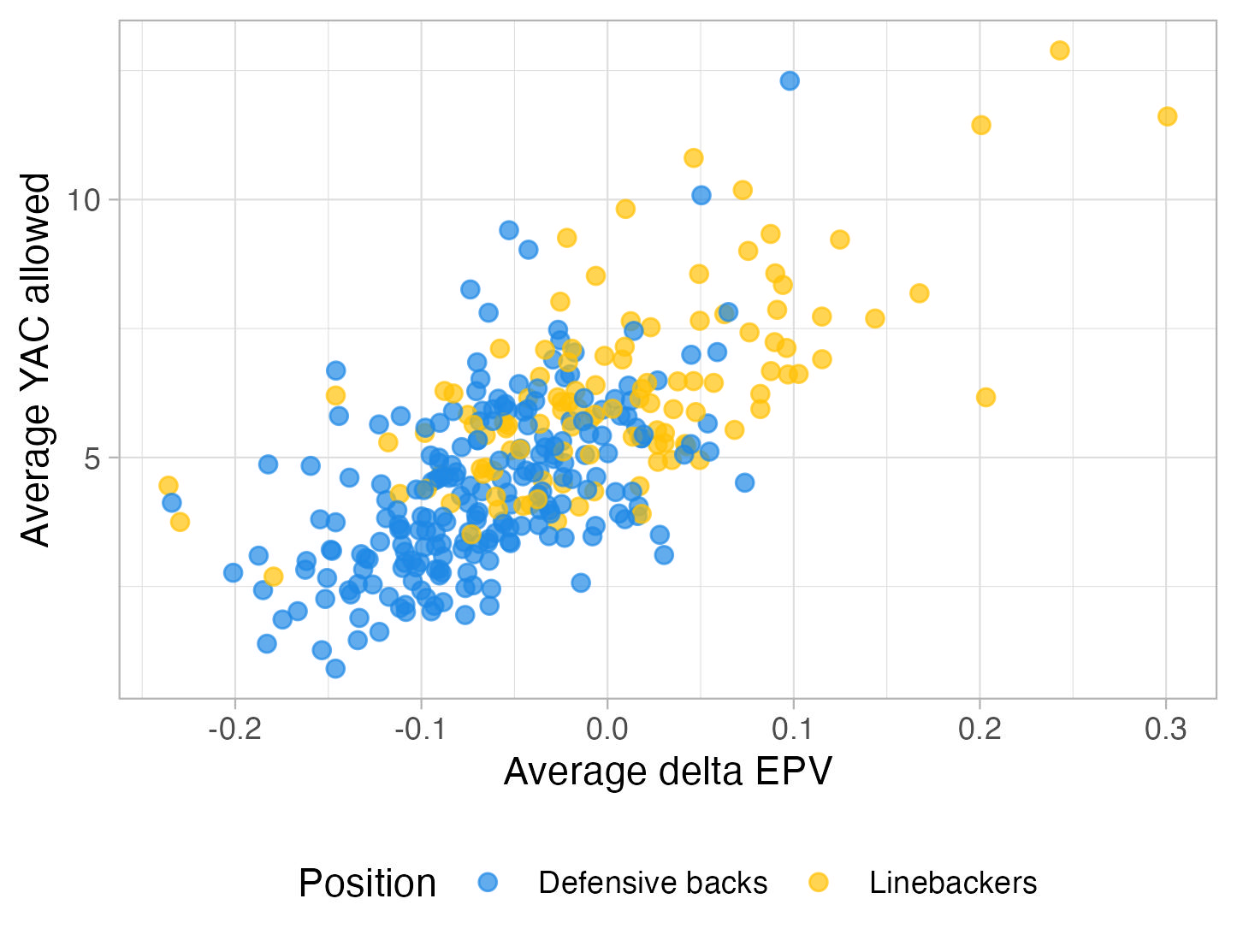}
    \caption{\amend{Relationship between the average YAC allowed (y-axis) and average $\mathbb{E}[\delta_{catch}]$ (average delta EPV, x-axis) across the 2018 NFL season for defenders with at least ten plays where they are the nearest defender to the receiver at the moment of catch. The overall correlation is moderately strong ($r = 0.65$) and there are similar levels of correlation within the two position groups: defensive backs (in blue, $r = 0.51$) and linebackers (in yellow, $r = 0.65$).}}
    \label{fig:ave-epv-yac-scatter}
\end{figure}

\amend{
Table \ref{tab:leaders} displays the top ten NFL defenders based on their accumulated $\mathbb{E}[\delta_{catch}]$ values across all plays where they are the closest to the receiver at the moment of catch. This leaderboard is dominated by cornerbacks (CB), with only two safeties (Von Bell and Derwin James). We notice that the top CB (Jaire Alexander) accumulates a total $\mathbb{E}[\delta_{catch}]$ of -7.37 across the 53 receptions where he is the nearest defender to the receiver. This means that Alexander's positioning and trajectory at the moment of catch is effectively seven points better than the ghost baseline across the entire 2018 NFL season. Since this aggregate is based on the change in expected points, it is measured in an interpretable scale which relates to actual game outcomes. In other words, Alexander's spatial positioning when the receiver catches the ball is worth approximately a single touchdown \amend{(plus extra point)} across the entire regular season. 

We also present in Table \ref{tab:leaders} the observed YAC allowed for these defenders for the plays in which they are the nearest defender at the moment of catch, since our EPV-based values do not actually account for the play outcome. We explore the relationship between our framework's values and the actual play outcome in terms of YAC in Figure \ref{fig:ave-epv-yac-scatter}. We observe a moderately strong positive relationship between the average $\mathbb{E}[\delta_{catch}]$ and the average YAC allowed for defenders that faced at least ten receptions as the nearest defender at the moment of catch. This indicates that players who tend to be more optimally positioned relative to the ghost baseline, thus more negative average $\mathbb{E}[\delta_{catch}]$ values, also allow fewer YAC on average. Figure \ref{fig:ave-epv-yac-scatter} also stresses a positional separation between defensive backs (cornerbacks and safeties) and linebackers. Linebackers are traditionally expected to be weaker in pass coverage relative to defensive backs, and we see that linebackers display more positive average $\delta$ values. From a player evaluation perspective, this motivates conditioning on the player's roster position in the 2D ghost RFCDE model. For now, a post-hoc  adjustment can be used for our results by simply taking the difference in the total or average values for players relative to their respective roster roles. We leave the consideration of roster position and other categorical variables for future work.}

\begin{table}
\caption{\amend{Top ten defensive players based on total accumulated $\mathbb{E}[\delta_{catch}]$ across the 2018 NFL season. The total and average $\mathbb{E}[\delta_{catch}]$ values are displayed (where lower values indicate better defensive positioning than ghost baseline), along with the total and average YAC allowed for the number of plays where they are the nearest defender to the receiver at the moment of catch.}}
\label{tab:leaders}
\centering
\begin{tabular}{llccccc}
\hline
Player's & Roster & Receptions & Total & Total & Average &  Average \\
Name & Position & Faced & $\mathbb{E}[\delta_{catch}]$ & YAC & $\mathbb{E}[\delta_{catch}]$ &  YAC \\
\hline
Jaire Alexander & CB & 53 & -7.37 & 128.46 & -0.14 & 2.42 \\ 
Adoree' Jackson & CB & 56 & -7.35 & 158.32 & -0.13 & 2.83 \\ 
A.J. Bouye & CB & 41 & -7.16 & 76.19 & -0.17 & 1.86 \\ 
Johnathan Joseph & CB & 50 & -6.49 & 152.47 & -0.13 & 3.05 \\ 
Malcolm Butler & CB & 48 & -5.87 & 161.48 & -0.12 & 3.36 \\ 
Vonn Bell & FS & 36 & -5.82 & 107.72 & -0.16 & 2.99 \\ 
Kendall Fuller & CB & 64 & -5.79 & 319.73 & -0.09 & 5.00 \\ 
Derwin James & SS & 40 & -5.77 & 232.20 & -0.14 & 5.80 \\ 
Jason McCourty & CB & 47 & -5.72 & 210.57 & -0.12 & 4.48 \\ 
Desmond Trufant & CB & 52 & -5.63 & 104.28 & -0.11 & 2.01 \\ 
\hline
\end{tabular}
\end{table}

\begin{figure}
    \centering
    \includegraphics[width = 0.6\textwidth]{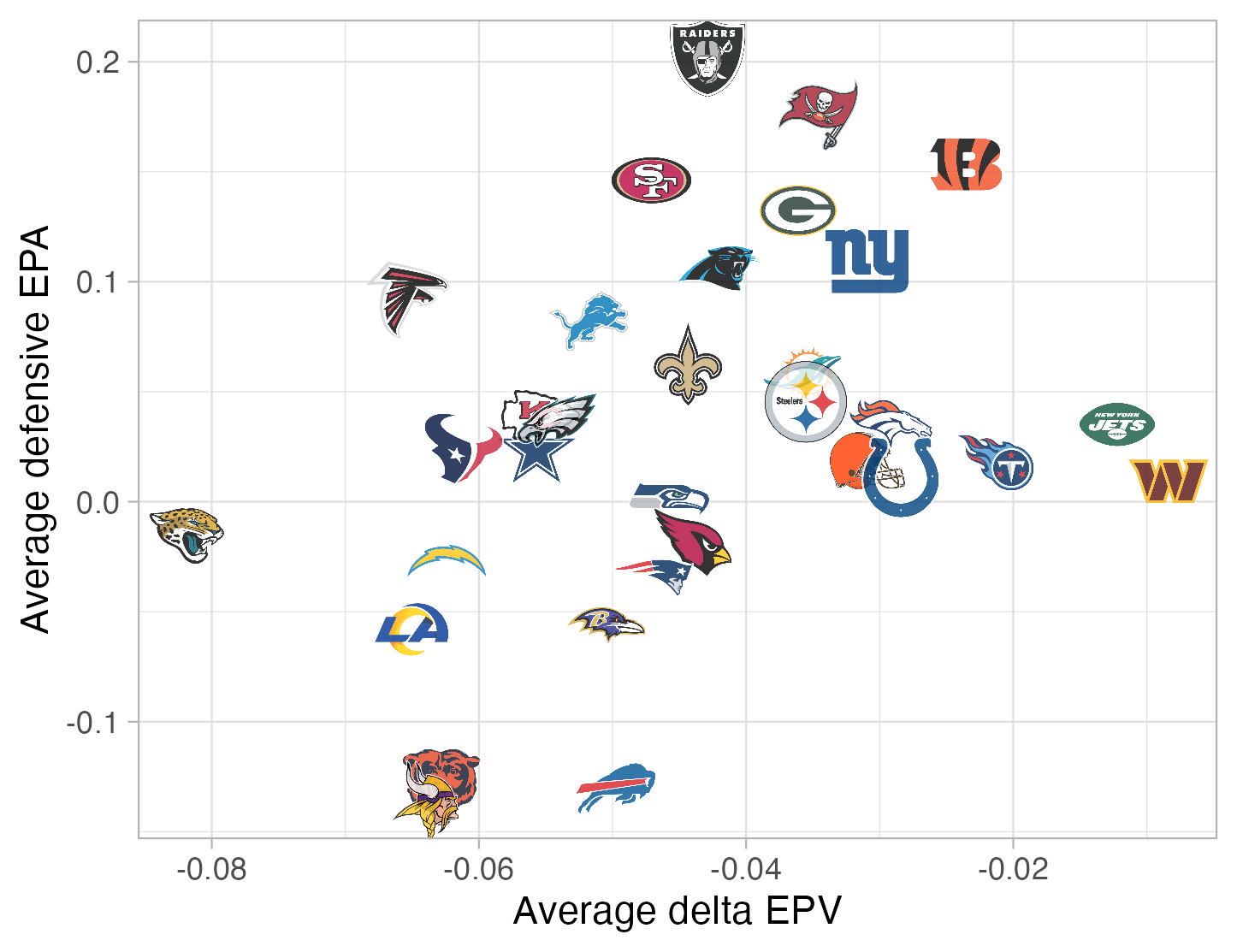}
    \caption{\amend{Scatterplot displaying relationship between team-level passing defensive EPA (y-axis) and $\mathbb{E}[\delta_{catch}]$ (average delta EPV, x-axis) across the 2018 NFL season.} Teams are displayed by their respective logos using the \texttt{nflplotR} package in \texttt{R} \citep{carl2024nflplotr}. Negative passing defensive EPA values indicate better team-level defensive performance.}
    \label{fig:team-plot}
\end{figure}

\amend{
Although we demonstrate our framework for evaluating the spatial positioning of defenders at only the moment of catch, we observe in Figure \ref{fig:team-plot} that our novel metrics (in terms of change in EPV) are associated with aggregate measures of team passing defense performance using expected points added (EPA) available from the \texttt{nflfastR} package in \texttt{R} \citep{carl2024nflfastr}. These pass defense EPA metrics are outcome-based, and include all pass attempts unlike our metrics which are only for passes that are complete. In Figure \ref{fig:team-plot}, we observe a moderate positive relationship ($r = 0.37$) between each team's average pass defensive EPA and our novel average $\mathbb{E}[\delta_{catch}]$ values across all receptions allowed by the team.} These results indicate that more optimal spatial positioning and defender trajectories at the moment of catch is related to better aggregate pass defense performance. For example, the Chicago Bears were considered one of the best defensive teams during the 2018 NFL season, with the best aggregate pass defense EPA relative to other teams. Our change in EPV metrics is in agreement as the Bears defenders on completed passing plays have the best total change in EPV relative to the ghost baseline. While this may sound unsurprising, our novel framework is the first approach to properly quantify and evaluate player positioning and trajectory information in American football.

\section{Discussion}
\label{sec:discussion}
In this work, we introduce a framework for evaluating the spatial positioning and trajectory of NFL defenders at the moment of catch relative to baseline ghost defenders. Our framework is the first public approach to evaluate the observed tracking data of players with proper consideration that hypothetical ghost players come from a distribution. We illustrate our framework in the context of modeling the nearest defender positioning at the moment of catch and estimate how much better or worse their positional information compares to ghosts with regards to the \amend{receiver's} value added through yards after catch. Using flexible random forests for conditional density estimation, our proposed framework provides estimates for (1) the distribution of a receiver's YAC which enables the estimation of the expected value of an interpretable utility function, and (2) the 2D distribution of hypothetical defender locations, providing full uncertainty quantification for baseline comparisons of defender positioning. This approach allows us to develop new metrics for player valuation using tracking data, and we demonstrate how such metrics are related to aggregate measures of performance.

We recognize that our framework and results are subject to various limitations. With regards to evaluating the spatial positioning of players, we cannot attribute the entire change in $EPV$ to a single player without recognizing that the scheme of the team's coaching staff may have played a large role. Additionally, we only \amend{consider evaluating} the nearest defender to the receiver at the moment of catch in our framework. \amend{First, this} definition of nearest defender may be viewed as undesirable, since by chance a defender that is covering a different receiver may be physically closer than the defender that is actually responsible for covering the receiver. There is opportunity to address this concern \amend{of coverage assignment} via a Hidden Markov Model approach, such as a basketball example demonstrated by \cite{franks2015characterizing}. 

\amend{Additionally, the nearest defender may not be the most ``important'' defender when it comes to limiting the yards gained after the catch. For example, a defender that is not located closest to the receiver but is downfield towards the receiver's target endzone (e.g., a safety between the receiver and endzone) may actually be the most important defender with regards to limiting the receiver's ability to score a TD after the catch. However, our model only considers covariate information at the moment of catch and fails to observe an improvement in the RFCDE performance beyond the considered set of features. One possible extension of our work with a richer dataset that includes all players on the field is to consider estimating the conditional density estimate of the ending yard line in Equation \ref{eq:epv-integral} with a flexible deep learning architecture through DeepCDE \citep{dalmasso2020conditional}. We also note that if tracking data for all players throughout a play were included in such an estimate, we could potentially identify the players with the most ``leverage'', i.e., changes in the $EPV$ value, based on perturbations in their tracking data features.} Furthermore, our framework imposes a simplified representation of American football, as we imagine that we can effectively move the nearest defender around without altering the positions of their teammates. Modeling a team's defense together, instead of just a single defender, is a challenging problem that could address this concern. However, we leave these problems and considerations for accounting for additional information (e.g., coverage type, receiver route type) for future work that can expand on our framework with more publicly-released data by the NFL over time.

Throughout this work, we emphasize that our main contribution is the framework in which we are evaluating the observed tracking data of players with an appropriate baseline. Our consideration for only considering a defender's positioning at the moment of catch with regards to receiver's YAC is relatively simple compared to the rest of play. Indeed, in order for the catch to be made, this means the defender has already ``failed'' in preventing the receiver from catching the football. \amend{There may be a tradeoff between the positioning of a defender with regards to limiting the catch probability versus the value gained after the catch.} We could extend our framework and attempt to evaluate \amend{the defender's tracking data} prior to completion, however this would require additional models for whether a receiver is targeted by the quarterback and catches the football \citep{yurko2020going, deshpande2020expected}. Moving beyond a single moment of time, we look forward to implementing our framework in full continuous-time and dealing with the challenges of modeling temporal ghost distributions to create new ways of measuring player performance in American football.

\section*{Acknowledgements}
The authors would like to thank Mike Lopez and the NFL Big Data Bowl 2021 for releasing publicly-available player tracking data, as well as Sam Ventura, Nic Dalmasso, the attendees and organizers of the 2023 New England Symposium on Statistics in Sports, and participants at the 2024 CMSAC Football Analytics Workshop for their valuable feedback.

\bibliographystyle{apalike}
\bibliography{bibliography}

\newpage
\section*{Supplementary Material}
\beginsupplement

\begin{figure}[!htb]
    \centering
    \includegraphics[scale = .15]{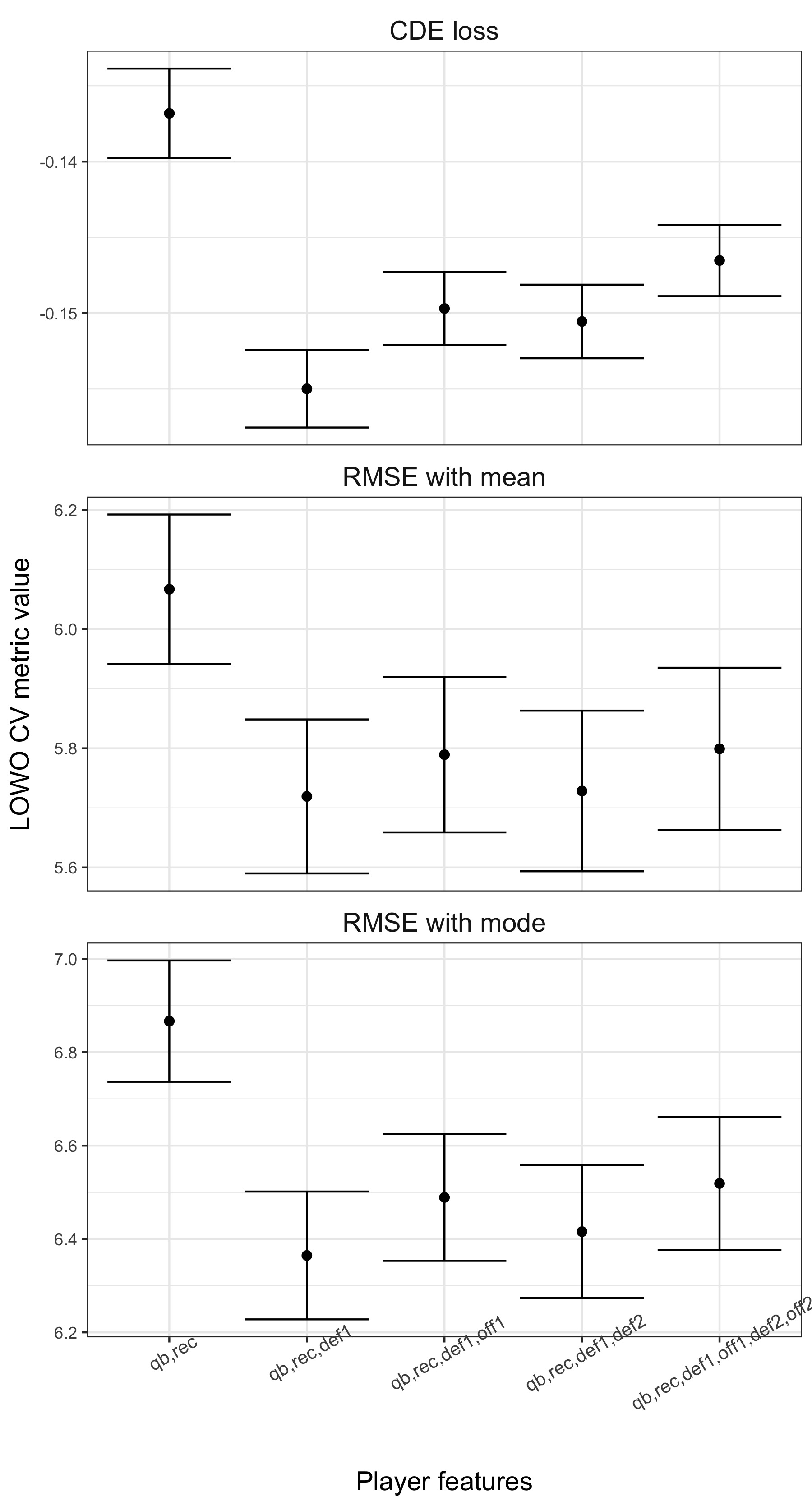}
    \caption{Comparison of leave-one-week-out cross validation performance (y-axis) for YAC RFCDE based on set of player features accounted for (x-axis) for three different metrics (in order): CDE loss, RMSE using RFCDE mean, and RMSE using RFCDE mode. Mean values are displayed as points with intervals for plus/minus one standard error.}
    \label{fig:suppl-yac-cv}
\end{figure}

\begin{figure}[!htb]
    \centering
    \includegraphics[scale = .15]{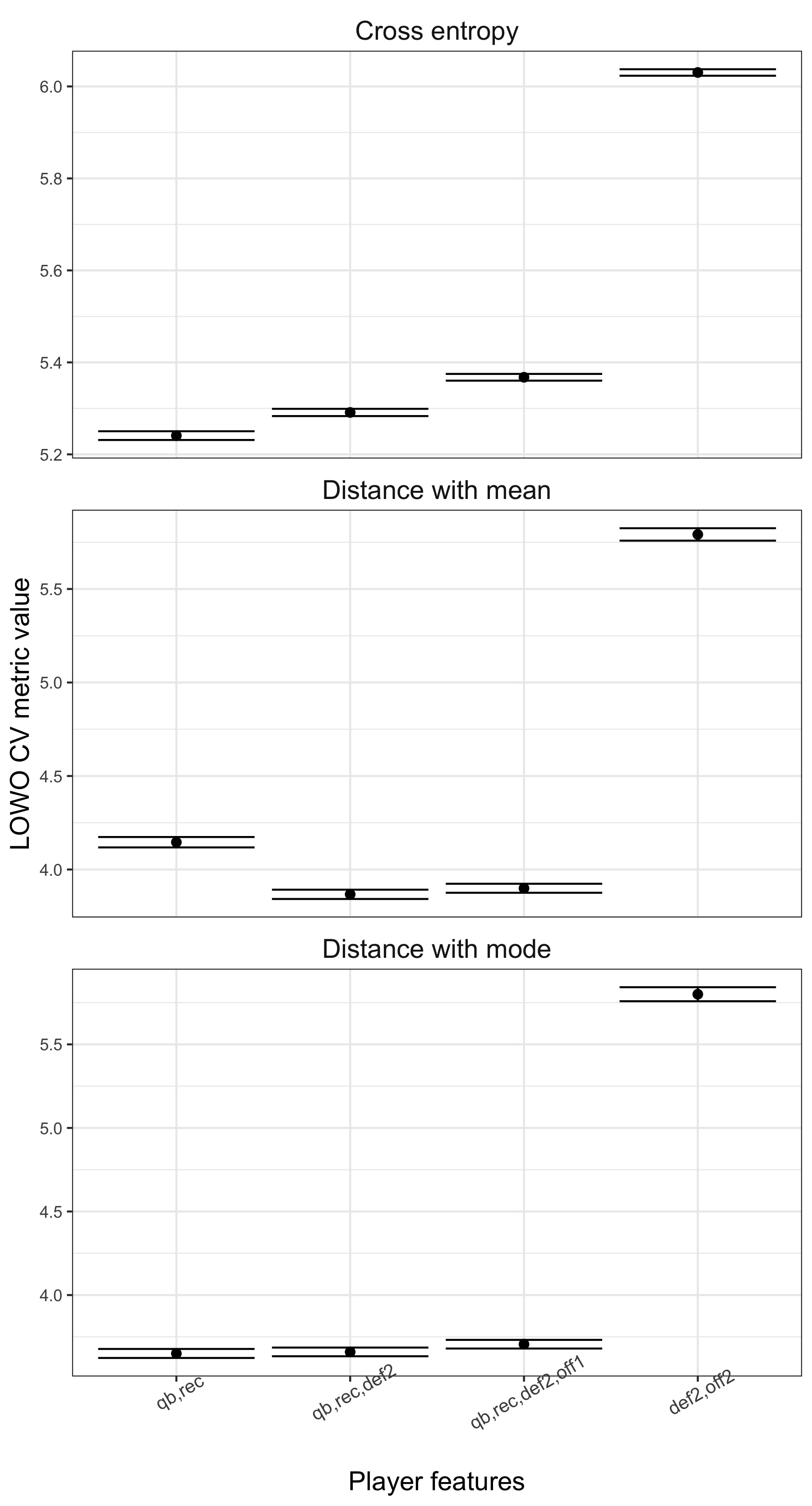}
    \caption{Comparison of leave-one-week-out cross validation performance (y-axis) for nearest defender 2D RFCDE based on set of player features accounted for (x-axis) for three different metrics (in order): cross entropy, distance from observed location using RFCDE mean, and distance from observed location using RFCDE mode. Mean values are displayed as points with intervals for plus/minus one standard error.}
    \label{fig:suppl-ghost-cv}
\end{figure}

\begin{figure}[!htb]
    \centering
    \includegraphics[scale = .15]{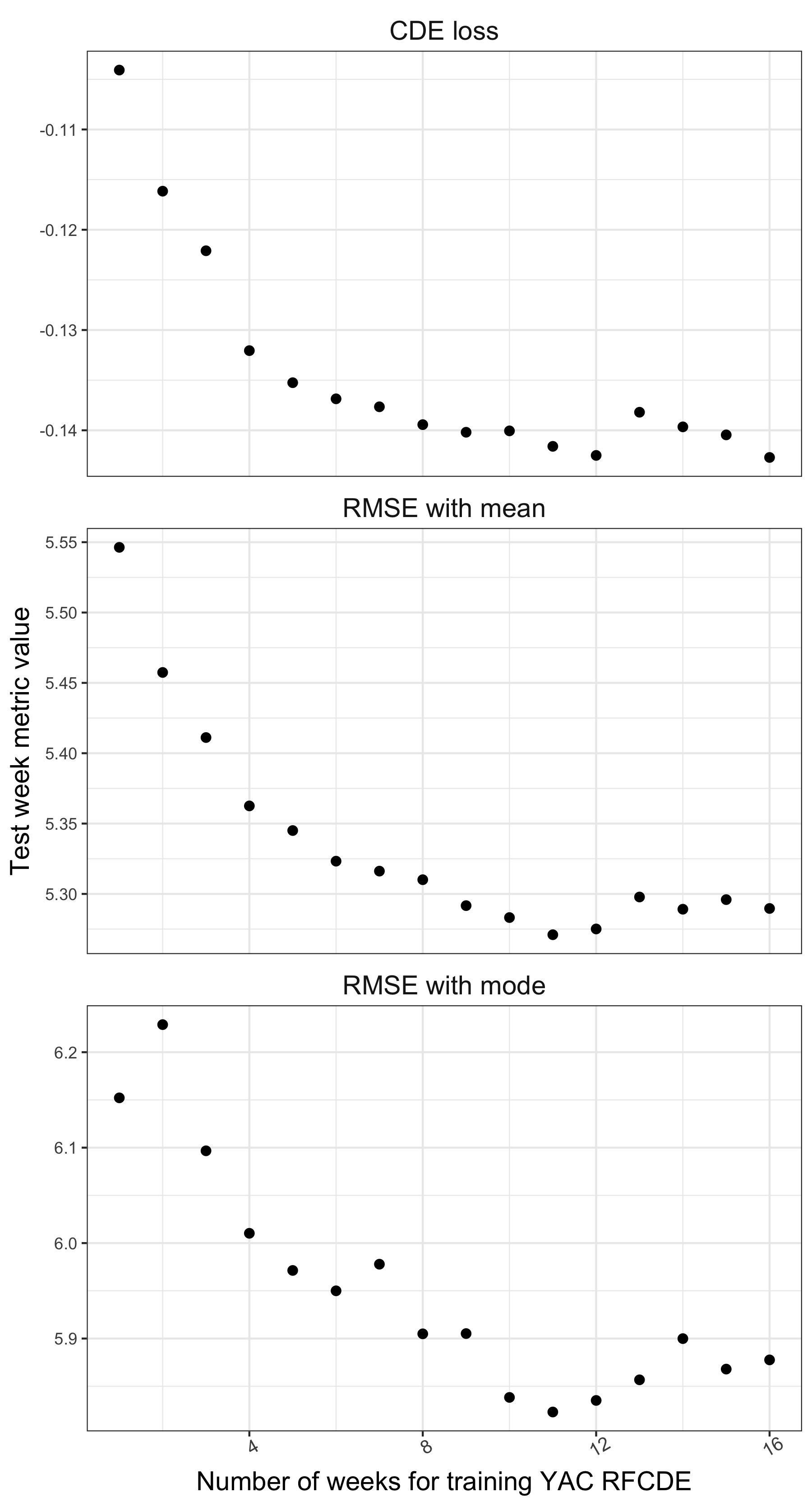}
    \caption{Sensitivity analysis to observe the performance of the YAC RFCDE as a function of the amount of training data. We evaluated the YAC RFCDE performance on the final Week 17 using CDE loss, RMSE using RFCDE mean, and RMSE using RFCDE mode as a function of the number of weeks of training data (x-axis) ranging from 1 to 16. }
    \label{fig:suppl-yac-week-metrics}
\end{figure}

\begin{figure}[!htb]
    \centering
    \includegraphics[scale = .15]{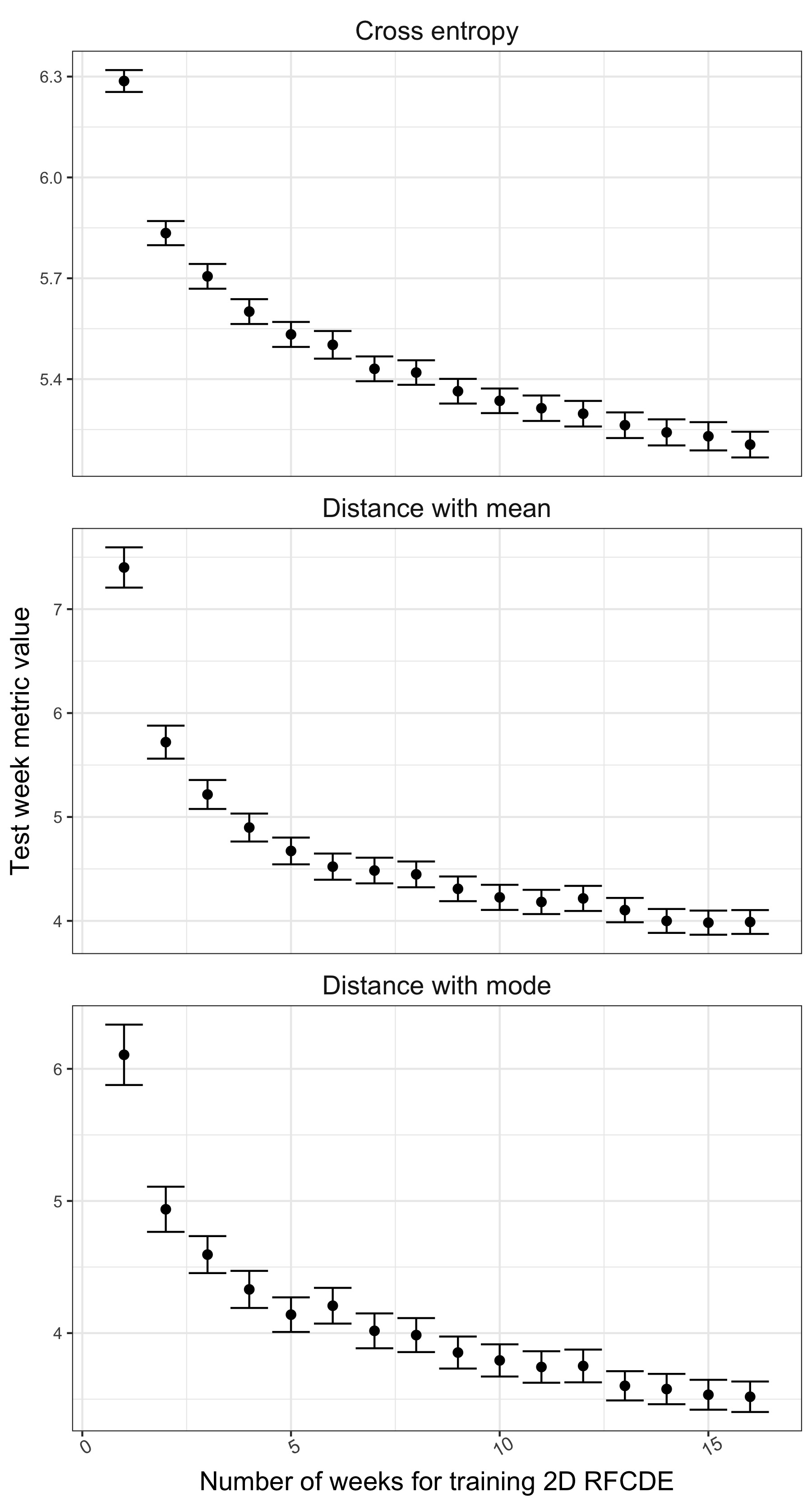}
    \caption{Sensitivity analysis to observe the performance of the 2D locaation RFCDE as a function of the amount of training data. We evaluated the 2D location RFCDE performance on the final Week 17 using cross entropy, distance from observed location using RFCDE mean, and distance from observed location using RFCDE mode as a function of the number of weeks of training data (x-axis) ranging from 1 to 16. Mean values are displayed as points with intervals for plus/minus one standard error.}
    \label{fig:suppl-ghost-week-metrics}
\end{figure}

\begin{figure}[!htb]
    \centering
    \includegraphics[width = \textwidth]{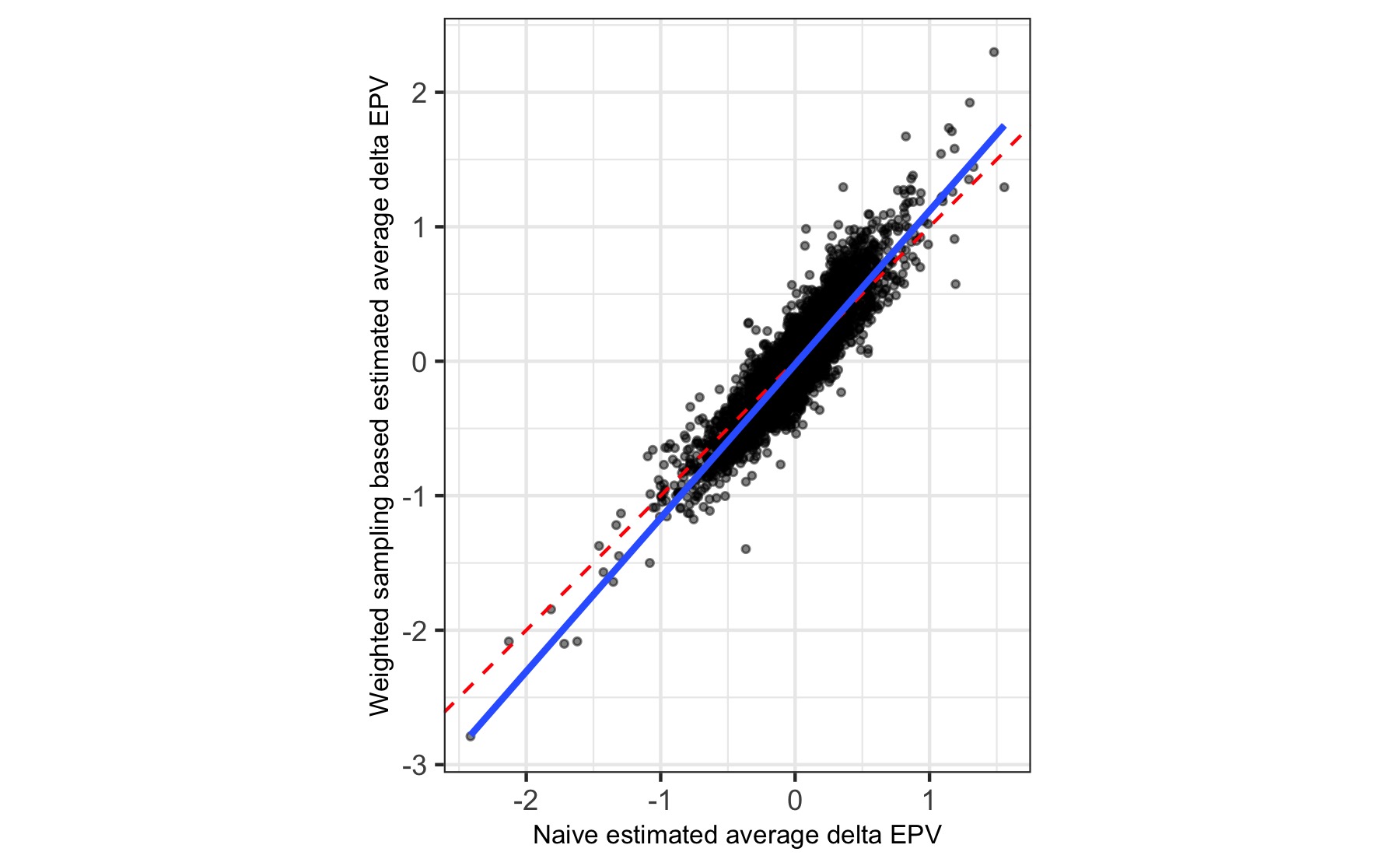}
    \caption{Scatterplot displaying the comparison between the estimates for the play-level $\mathbb{E}[\delta_{catch}]$ values where the observed defender's trajectory (speed, direction, orientation) are used (x-axis) versus our weighted sampling strategy (y-axis) that tries to account for the relationship between the player's location and their trajectory relative to the receiver. The identity $y=x$ (red dashed line) and linear regression fit (blue solid line) are displayed. We observe a strong positive relationship between the two different approaches (correlation $\approx 0.90$) indicating that the rough ordering of value will be similar, yet our weighted sampling approach can lead to swings in estimated value that are close to one point.}
    \label{fig:suppl-delta-epv-comparison}
\end{figure}

\end{document}